\documentclass[twocolumn,prb,showpacs,graphicx,superscriptaddress]{revtex4}
\usepackage{graphicx}
\usepackage{dcolumn}
\usepackage{bm}
\usepackage[cmex10]{amsmath}
\usepackage[utf8]{inputenc}
\usepackage{float}
\usepackage{amssymb}
\usepackage[T1]{fontenc}
\usepackage{natbib,hyperref}

\usepackage{epsfig}

\begin{document}

\preprint{APS/123-QED}

\title{Rectification of radio frequency current in giant magnetoresistance spin valve}

\author{Sławomir Ziętek}
\email{zietek@agh.edu.pl}
\affiliation{AGH University of Science and Technology, Department of Electronics, Al. Mickiewicza 30, 30-059 Krak\'{o}w, Poland}

\author{Piotr Ogrodnik}
\email{piotrogr@if.pw.edu.pl}
\affiliation{AGH University of Science and Technology, Department of Electronics, Al. Mickiewicza 30, 30-059 Krak\'{o}w, Poland}
\affiliation{Warsaw University of Technology, Faculty of Physics, , ul. Koszykowa 75, 00-662 Warszawa, Poland}

\author{Marek Frankowski}
\email{mfrankow@agh.edu.pl}
\affiliation{AGH University of Science and Technology, Department of Electronics, Al. Mickiewicza 30, 30-059 Krak\'{o}w, Poland}

\author{Jakub Chęci\'{n}ski}
\affiliation{AGH University of Science and Technology, Department of Electronics, Al. Mickiewicza 30, 30-059 Krak\'{o}w, Poland}
\affiliation{AGH University of Science and Technology, Faculty of Physics and Applied Computer Science, Al. Mickiewicza 30, 30-059 Kraków, Poland}

\author{Piotr Wiśniowski}
\affiliation{AGH University of Science and Technology, Department of Electronics, Al. Mickiewicza 30, 30-059 Krak\'{o}w, Poland}

\author{Witold Skowro\'{n}ski}
\affiliation{AGH University of Science and Technology, Department of Electronics, Al. Mickiewicza 30, 30-059 Krak\'{o}w, Poland}

\author{Jerzy Wrona}
\affiliation{AGH University of Science and Technology, Department of Electronics, Al. Mickiewicza 30, 30-059 Krak\'{o}w, Poland}
\affiliation{Singulus Technologies, Kahl am Main, 63796, Germany}

\author{Tomasz Stobiecki}
\affiliation{AGH University of Science and Technology, Department of Electronics, Al. Mickiewicza 30, 30-059 Krak\'{o}w, Poland}

\author{Antoni Żywczak}
\affiliation{AGH University of Science and Technology, Academic Centre of Materials and Nanotechnology, Al. Mickiewicza 30, 30-059 Kraków, Poland}

\author{J\'{o}zef Barna\'{s}}
\affiliation{Adam Mickiewicz University, Faculty of Physics, ul. Umultowska 85, 61-614 Poznań, Poland}
\affiliation{Institute of Molecular Physics, Polish Academy of Sciences, ul. Smoluchowskiego 17, 60-179 Poznań, Poland}


\begin {abstract}
We report on a highly efficient spin diode effect in an exchange-biased spin-valve giant magnetoresistance (GMR) strips. In such multilayer structures, symmetry of the current distribution along the vertical direction is broken and, as a result, a non-compensated Oersted field acting on the magnetic free layer appears. This field, in turn, is a driving force of magnetization precessions. Due to the GMR effect, resistance of the strip oscillates following the magnetization dynamics. This leads to rectification of the applied radio frequency current and induces a direct current voltage $V_{DC}$. We present a theoretical description of this phenomenon and calculate the spin diode signal, $V_{DC}$, as a function of frequency, external magnetic field, and angle at which the external field is applied. A satisfactory  quantitative agreement between theoretical predictions and experimental data has been achieved. Finally, we show that the spin diode signal in GMR devices is significantly stronger than in the anisotropic magnetoresistance permalloy-based devices.

\end{abstract}

\maketitle

\section{Introduction}

Radio Frequency (RF) devices have  been  of significant interest for a long time due to their multiple applications, e.g. in wireless telecommunication, fast electronics or radar technologies. Due to certain limitations of the semiconductor technology, new materials and phenomena that could be applied in microwave devices are highly desired.\cite{prokopenko2013spin,kima2012spin} Ferromagnetic Resonance (FMR) in magnetic multilayer systems, which is typically probed at RF regime, gives a chance to create new microwave nanodevices such as filters, rectifiers, oscillators, phase shifters or delay lines.\cite{prokopenko2013spin,sattler2010handbook}

An AC current passing through a magnetic structure can entail an oscillation of the magnetization due to the spin transfer torque, field torque or spin-orbit torque.~\cite{yamaguchi2007rectification,harder2011analysis,mecking2007microwave,kurebayashi2014antidamping} This oscillation, in turn, results in the variation of the resistance due to the magnetoresistance effect. The oscillating resistance mixed with the AC current gives rise to a DC voltage component, and this rectification is called a spin diode effect,\cite{tulapurkar2005spin} which has been widely investigated in nanostructured Magnetic Tunnel Junctions (MTJs)~\cite{tulapurkar2005spin,sankey2006spin,sankey2007measurement,skowronski2013influence,skowronski2014spin,kubota2007quantitative}
as well as in different thin-film systems based on Py,~\cite{yamaguchi2007rectification,gui2005resonances,costache2006large,gui2007realization} Fe,~\cite{hui2008electric} and other materials and compounds.~\cite{goennenwein2007electrically,wirthmann2008broadband,costache2006electrical,saitoh2006conversion,
mosendz2010quantifying,mosendz2010detection,saraiva2010dipolar,kajiwara2010transmission,
sandweg2010enhancement,liu2011spin,azevedo2011spin,kurebayashi2014antidamping}

Extensive theoretical research has been carried out in order to examine the physical foundations of the phenomena involved in RF devices.\cite{harder2011analysis,yamaguchi2007rectification,mecking2007microwave} It has been shown that a proper analysis of the FMR-generated DC voltage requires distinguishing between different mechanisms contributing to the symmetrical and antisymmetrical components of the signal.\cite{harder2011analysis} The role of the relative phase difference between RF electric and magnetic fields has also been discussed in literature, as summarized by Harder et al. in Ref.~\onlinecite{harder2011analysis}.
Although experimental data reported by several groups are consistent, there are some discrepancies in theoretical descriptions of the physical origin of the DC voltage V$_{DC}$.

Most of the recent studies on spin diode effect were focused either on devices exhibiting Anisotropic Magnetoresistance (AMR) effect or on MTJs with the Tunneling Magnetoresistance (TMR). The spin diode signal in AMR devices is usually very weak due to a low MR ratio. In MTJs, on the other hand, a much stronger diode signal is usually observed. Operation of the MTJ-based devices, however, is limited to voltages below the corresponding breakdown values. In this paper we report on a spin diode effect in an exchange-biased Spin Valve Giant Magnetoresistance (SV-GMR) strip, which is an alternative to AMR- or MTJ-based devices. Very recently, a similar SV-GMR strip has been investigated by Kleinlein et al.~\cite{kleinlein2014using} The authors, however, focused their attention mainly on its enhanced detection sensitivity when compared to an AMR-based device.

In section \ref{sec:theory}, we provide a theoretical description of the DC voltage and the electrically detected FMR spectra. In section \ref{sec:experimental}, we describe the preparation process of the investigated multilayer strip, the experimental setup used to detect a spin diode signal and the origin of the Oersted field  affecting the free layer dynamics. Obtained results, including angular and frequency dependences of the signal, are presented in section \ref{sec:results}, where we also discuss the nature of the magnetization oscillations responsible for time-dependent resistance changes, as well as the physical origin of the DC voltage signal. The experimental data are compared with theoretical predictions for angular and frequency dependences of the spin diode signal. Finally, section \ref{sec:summary} contains a short summary and conclusions.

\section{Theory\label{sec:theory}}
\subsection{General background on  $V_{DC}$}

\begin{figure}[!htbp]
\centering
\includegraphics[width=8.6cm]{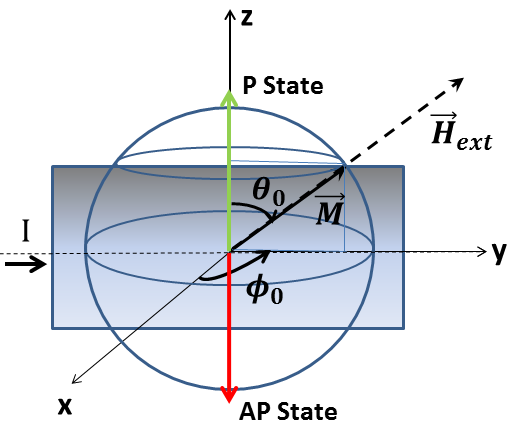}
\caption{  Top view of the SV-GMR structure under consideration: magnetization of the RL is oriented along the axis $z$, while that of the FL is oriented along the external in-plane ($y-z$) magnetic field direction. The polar ($\theta_0$) and azimuthal ($\phi_0$) angles, which determine orientation of the FL magnetization in equilibrium situation are also indicated. Parallel (P) state corresponds to
$\theta_0=0$, while the antiparallel (AP) configuration to $\theta_0=\pi$. The AC current flowing through the structure is oriented along the axis $y$. In this figure, magnetization is aligned along the external field direction since the uniaxial anisotropy field in our experiment is negligibly small (see section \ref{ssec:fabrication}).) }
\label{GMRGEOM}
\end{figure}

In this section we present some general theoretical background on the spin diode effect, based on the description by Nozaki et al. \cite{nozaki2012electric} for magnetoresistive tunnel
junctions. This description is then adapted to the experimental SV-GMR system. For this purpose, we define a coordinate system with respect to the sample orientation, as presented in Fig.\ref{GMRGEOM}. Consecutive layers of the structure lie in the $y-z$ plane, and
we assume that the magnetization of the Reference Layer (RL) is pinned along the $z$ axis and does not affect the dynamics of the Free Layer (FL).
The polar angle $\theta$ is defined as the angle between the magnetization $\vec{M}$ of the FL and magnetization of the RL. This angle also determines the resistance of the SV-GMR according to the formula,
\begin{equation}
 R(\theta) = R_{P} + \frac{\Delta R}{2}(1-\cos\theta),
\label{RGMR}
\end{equation}
where $\Delta R=R_{AP}-R_{P}$, while $R_P$ and $R_{AP}$ denote resistances in the Parallel (P) and  Antiparallel (AP) state, respectively. 

A well-established fact is that the application of a radio-frequency voltage, $V(t)=V\cos(\omega t)=\mathcal{R}\{Ve^{i\omega t}\}$, to a magnetoresistive strip generates a
time-dependent driving force (due to spin torque, \cite{sankey2007measurement,yamaguchi2007rectification} Oersted field, \cite{thiaville2008electrical} or anisotropy field\cite{nozaki2012electric}) which may generate oscillations of the magnetization of the FL. These oscillations result in small changes of the resistance: $\delta R(t) = \bar{\delta R} \cos(\omega t+\beta)=\mathcal{R}\{\bar{\delta R} e^{i(\omega t+\beta)}\}$. Here $\beta$ is a phase shift between the time-dependent current and the resistance, while $\bar{\delta R}$ is the amplitude (real) of the resistance change. Since the AC current flowing through the sample depends mainly on the resistance in the corresponding  stationary point described by $\theta_0$ and $\phi_0$,  it can be approximated by $I(t) = \frac{V\cos(\omega t)}{R(\theta_0)}$, and thus the applied AC voltage and AC current are in-phase. Note that the stationary point is determined by the magnetic energy in the absence of the AC signal.

The output DC voltage $V_{DC}$ is determined by the product of the time-dependent resistance and the current. Apart from the DC voltage, this product also includes the AC voltage $V_{AC}$ of doubled frequency. Thus, we can write:
\begin{equation}
V_{out} = V_{DC}+V_{AC} = \frac{1}{R(\theta_0)} \mathcal{R}\{Ve^{i\omega t}\}\mathcal{R}\{\bar{\delta R} e^{i(\omega t+\beta)}\}.
\label{eq:average}
\end{equation}
From the above equation one easily finds
\begin{equation}
V_{DC}=\frac{V}{2}\frac{\bar{\delta R}}{R(\theta_0)}\cos(\beta).
\label{eq:simple}
\end{equation}
The voltage $V_{DC}$ can be detected for instance in a FMR dynamics experiment.\cite{nozaki2012electric} The point is that the phase shift $\beta$ and the amplitude of the resistance change $\bar{\delta R}$ are not constant, but generally depend on the frequency. Therefore, the simple formula (\ref{eq:simple}) for $V_{DC}$  is rather useless for interpretation of experimental data, and one needs to derive a more general expression for $V_{DC}$. To do this, we rewrite the expression for
the $V_{DC}$ voltage in the form in which the phase shift is not extracted explicitly, but is included in the resistance change, $\bar{\delta R}e^{i\beta}\equiv \delta R$,
\begin{equation}
V_{DC} = \eta \frac{V}{2}\frac{\mathcal{R}\{\delta R\}}{R(\theta_0)},
\label{vdcgen}
\end{equation}
where we introduced an additional phenomenological factor $\eta$ which originates from both parasitic impedances of measurement setup and the impedance of the sample. The parameter $\eta$ will be treated as a free parameter, and will allow us to compare the experimental and theoretical results quantitatively.\cite{nozaki2012electric} It is important to note that $\eta$ influences neither shape of the resonance spectra nor their linewidths and the resonant frequencies, but instead it acts rather as a scaling factor which modifies only the absolute values of the spectrum amplitude.

To determine the $V_{DC}$ signal, one needs to determine the resistance change, $\delta{R}$ in Eq.(\ref{vdcgen}) around the stationary point $\theta_0$ for a given AC voltage. Thus, using Eq.(\ref{RGMR}) and calculating the first derivative at $\theta_0$, one finds
\begin{equation}
\delta R = \frac{\Delta R}{2} \sin \theta_0 \delta \theta ,
\end{equation}
where $\delta \theta $ is a change in the polar angle due to the magnetization dynamics.
Combining this with  Eq.(\ref{vdcgen}) allows us to write the DC voltage signal in the form
\begin{equation}
V_{DC} = \eta \frac{V}{4}\frac{\Delta R}{R(\theta_0)}\sin\theta_0 \mathcal{R}\{\delta \theta\}
\label{vdcgmr}
\end{equation}
What we need now is to calculate the angle change $\delta \theta$ that can be derived from the Landau-Lifschitz-Gilbert (LLG) equation.

\maxdeadcycles=10000

\subsection{ FMR resonance theory of $V_{DC}$}

To calculate $\delta\theta$ we use the LLG equation in the macrospin approximation,
\begin{equation}
\frac{d\vec{M}}{dt}=  -\gamma_e\vec{M} \times \vec{H}_{\rm eff} +\frac{\alpha}{M_S} \vec{M} \times \frac{d\vec{M}}{dt} ,
\label{eq:LLG_wek}
\end{equation}
where $\gamma_e = \frac{g\mu_B}{\hbar}$ is the gyromagnetic ratio with spectroscopic splitting factor $g$ = $2.1$ ($\gamma_e>0$), $M_S$ denotes the saturation magnetization of the FL, $\vec{H}_{\rm eff}$ stands for the effective magnetic field which can be described by the corresponding magnetic energy density $U$, $\vec{H}_{\rm eff}=-\partial U/\partial \vec{M}$. In the SV-GMR structure under consideration, the driving force for FMR originates from the time-dependent Oersted field associated with the AC current flowing along the strip. This assumption will be discussed in more detail in the next section, where it will be shown that it is sufficient to describe experimental results. Generally, the Oersted field cannot be written as a gradient of a potential energy due to its rotational character, $\nabla \times \vec{H}_{Oe} \neq 0$.\cite{griffiths2013introduction} In our case, however,  we take into account the Oersted field only in the FL. This field is uniform and oriented along the $z$ axis, $\vec{H}_{Oe,z}\equiv H_{Oe}$, so that the associated energy is simply the Zeeman energy which can be included in $U$.

By introducing unit vectors
$\hat{e}_\theta$ and $\hat{e}_\phi$ associated with the spherical coordinates,\cite{weisstein2004spherical} one can rewrite the LLG equation,   Eq.(\ref{eq:LLG_wek}), as the following differential equation:
\begin{eqnarray}
(\sin\theta \;\dot{\phi} \hat{e}_\phi + \dot{\theta}\hat{e}_{\theta})+\alpha\sin\theta \;\dot{\phi} \hat{e}_{\theta}-\alpha\dot{\theta}\hat{e}_\phi \nonumber \\
= \gamma_e \frac{\partial U}{\partial \theta}\hat{e}_\phi - \frac{\gamma_e}{\sin \theta} \frac{\partial U}{\partial \phi}\hat{e}_{\theta}.
\label{eq:LLG_form3}
\end{eqnarray}
Since the magnetization of the FL is driven by a periodic Oersted field, one can assume that both spherical angles oscillate periodically with small amplitudes around the stationary point ($\theta_0,\phi_0$);
$\theta(t)=\theta_0 + \delta\theta e^{i\omega t}$ and
$\phi(t)=\phi_0 + \delta\phi e^{i\omega t}$. Note that $\delta\theta$ and $\delta\phi$ include possible phase shift between the magnetization oscillation and the driving AC current.
Equation  Eq.(\ref{eq:LLG_form3}) can be then rewritten as two coupled equations for $\delta\theta$ and $\delta\phi$,
\begin{equation}
\left( \begin{array}{c}
i\omega\delta\theta \\
i\omega\delta\phi \end{array}
\right)
= \left( \begin{array}{c}
\frac{-\gamma_e}{1+\alpha^2}\left( \frac{1}{\sin\theta }\frac{\partial U}{\partial \phi} + \alpha \frac{\partial U}{\partial \theta} \right) \\
\frac{-\gamma_e}{1+\alpha^2}\left( -\frac{1}{\sin\theta }\frac{\partial U}{\partial \theta} + \frac{\alpha}{\sin^2\theta} \frac{\partial U}{\partial \phi} \right)
\end{array}
\right).\label{eq:postac_wyjsciowa}\end{equation}
Upon linearization with respect to small deviations $\delta\theta$ and $\delta\phi$ from the stationary point ($\theta_0, \phi_0$), the LLG equation (\ref{eq:postac_wyjsciowa}) takes the form
\begin{widetext}
$$
\left( \begin{array}{cc}
\left[ i\omega (1 + \alpha^2) + \frac{\gamma_e}{\sin\theta}B + \frac{\cos\theta}{\sin^2\theta}A+\alpha\gamma_e D   \right]
 &
 \gamma_e \left[
 \frac{1}{\sin\theta}C + \alpha B \right]
\\
\gamma_e \left[ \frac{\cos\theta}{\sin^2\theta}E - \frac{1}{\sin\theta}D -
2 \frac{\alpha\cos\theta}{\sin^3\theta}A + \frac{\alpha}{\sin^2\theta}B \right]
 & \left[ i\omega (1 + \alpha^2) - \gamma_e (B\frac{1}{\sin\theta} - \alpha
   C \frac{1}{\sin^2\theta})\right]
\end{array}
\right)
\left( \begin{array}{c}
\delta\theta \\
\delta\phi \end{array}
\right) = $$
\begin{equation}
= \left( \begin{array}{c}
-\gamma_e\left[ H\frac{1}{\sin\theta} + \alpha M \right]  \\
\gamma_e \left[ \frac{1}{\sin\theta}M - \alpha H \frac{1}{\sin^2 \theta} \right]
\end{array}
\right) H_{Oe}e^{i \psi},
\label{eq:postac_uproszczona}
\end{equation}
\end{widetext}
where we have introduced parameters denoting first and second derivatives of the magnetic energy: $A\equiv\frac{\partial U}{\partial \phi},
B\equiv \frac{\partial^2 U}{\partial \phi \partial\theta}, C\equiv\frac{\partial^2 U}{\partial\phi^2},D\equiv \frac{\partial^2 U}{\partial\theta^2},
E\equiv \frac{\partial U}{\partial \theta}, H\equiv \frac{\partial^2 U}{\partial\phi\partial H_{Oe}}, M\equiv \frac{\partial^2 U}{\partial\theta\partial H_{Oe}}$.
The introduced phase factor $\psi$ stands for the phase shift between the electric and the magnetic fields (or equivalently between the AC current and AC Oersted field).~\cite{harder2011analysis}

Equation~(\ref{eq:postac_uproszczona}) has the general form
$ \hat{A}\hat{X} =\hat{Y}$, where $ \hat{A}$ denotes the matrix on the left side of Eq.~(\ref{eq:postac_uproszczona}), $\hat{X}$ is the vector composed of $\delta\theta$ and $\delta\phi$ and $\hat{Y}$ is the right side of Eq.~(\ref{eq:postac_uproszczona}).
This equation can be solved by multiplying its both sides by the inverse matrix $\hat{A}^{-1}$. After some algebra, one  finds $\hat{A}^{-1}$
in the explicit form:
\begin{widetext}
\begin{equation}
\hat{A}^{-1}= \frac{-1}{\Gamma(\omega^2 - \omega^2_0 - i\omega\sigma)}
\left( \begin{array}{cc}
i (1 + \alpha^2)\omega +  \frac{-\gamma_e}{\sin\theta} ( B - C \frac{\alpha}{\sin\theta})
 & \gamma_e (-B \alpha - C \frac{1}{\sin\theta}) \\
\frac{\gamma_e}{\sin\theta} (D - E\frac{\cos\theta}{\sin\theta} - B \frac{\alpha}{\sin\theta} + 2 A \alpha\frac{\cos\theta}{\sin^2\theta})
 &
i (1 + \alpha^2) \omega - \gamma_e(A \frac{\cos\theta}{\sin^2\theta} - D\alpha - B \frac{1}{\sin\theta})
\end{array}
\right).
\label{eq:last_2_var}
\end{equation}
\end{widetext}
In the above equation $\Gamma$ is defined as  $\Gamma\equiv (1+\alpha^2)^2$, while the square of the angular resonance frequency and the corresponding linewidth are given respectively by the following formulas:
\\
\begin{widetext}
\begin{equation}
\omega^2_0\equiv
\frac{\gamma_e^2}{\Gamma}\frac{1}{\sin\theta} \left[ \frac{1+\alpha^2}{\sin\theta} (CD - B^2) +  \frac{\cos\theta}{\sin\theta} \left(-EB\alpha + \frac{1}{\sin\theta}(2\alpha^2 AB + AB - EC) +
\frac{\alpha}{\sin^2\theta}AC\right) \right],
\label{eq:resfreq}
\end{equation}
and
\begin{equation}
\sigma \equiv \frac{\gamma_e}{(1+\alpha^2)} \left[ \alpha D - \frac{\cos\theta}{\sin^2\theta} A + \frac{\alpha}{\sin^2\theta} C \right].
\label{eq:sigma}
\end{equation}
\end{widetext}

Taking into account Eqs~(\ref{eq:postac_uproszczona}) and (\ref{eq:last_2_var}) one can easily find the solutions for $\delta\theta$ and $\delta\phi$.  
Then, the real part of the solution for $\delta\theta$ can be introduced into Eq.(\ref{vdcgen}), which leads to the final result for the $V_{DC}$ signal originating from the Oersted field
for an arbitrary form of the magnetic energy:
\begin{widetext}
\begin{equation}
V_{DC} = \eta\frac{I H_{Oe}\Delta R\sin\theta}{4}\frac{\gamma_e(1+\alpha^2)}{\Gamma((\omega^2 - \omega^2_0)^2 + \omega^2\sigma^2)}
 \left[ \cos\psi\; Z_1- \sin\psi\; Z_2\right]
\label{eq:vdcGMR}
\end{equation}
where
\begin{equation}
Z_1 = -\sigma \omega^2 (M \alpha +  H \csc\theta) - \gamma_e\csc^2\theta\; (\omega^2 - \omega_ 0 ^2) (BH-CM)
\end{equation}
\begin{equation}
Z_2 = \omega(\omega^2- \omega_0 ^2) (M \alpha +  H \csc\theta) - \gamma_e\csc^2\theta \; \sigma\omega (BH-CM)
\end{equation}
\end{widetext}
and $I$ denotes the amplitude of the AC current flowing through the sample at a given angle $\theta_0$. The phase shift $\psi$ has a significant influence on the shape
of the FMR spectra and, for a given form of energy $U$, may change their character from antisymmetrical to symmetrical and vice versa.

To complete this part, we note that the factor $\left[ \cos\psi Z_1- \sin\psi Z_2\right]$ in Eq.(\ref{eq:vdcGMR}) can be written in the form consistent with
Eq.(\ref{eq:simple}). Writing $Z_1$ and $Z_2$ as $Z_1/(Z_1^2+Z_2^2)=\cos\Phi$ and $Z_2/(Z_1^2+Z_2^2)=\sin\Phi$,
one can rewrite this factor in the form
$\left[ \cos\psi Z_1- \sin\psi Z_2\right]=(Z_1^2+Z_2^2)[\cos\psi\cos\Phi-\sin\psi\sin\Phi]=(Z_1^2+Z_2^2)\cos(\psi+\Phi)$, i.e. in the form (\ref{eq:simple}) with
$\beta$ ($\beta=\psi+\Phi$) and the prefactor explicitly dependent on the frequency and other parameters of the model.

\section{Experimental\label{sec:experimental}}

\subsection{SV-GMR device: fabrication\label{ssec:fabrication}}

The material stack with the structure (nominal thicknesses in nm): Si / SiO$_2$ / Ta(3) / Ni$_{81}$Fe$_{19}$(1) / Pt$_{46}$Mn$_{54}$(18) / Co$_{90}$Fe$_{10}$(2) / Ru(0.85) / Co$_{90}$Fe$_{10}$(2.1) / Cu(2.1) / Co$_{90}$Fe$_{10}$(1) / Ni$_{81}$Fe$_{19}$ / (5) / Ru(0.5) /Cu(1) / Ta(3)  was deposited by TIMARIS magnetron sputtering system at Singulus AG. The easy axis of the magnetic layers was set by applying 100 Oe field during layers deposition. The Cu thickness (2.1 nm) was chosen to minimize the interlayer coupling between FL and RL. CoFe/NiFe composite was used to achieve a magnetically soft FL with large magnetoresistance sustained by the Cu/CoFe interface \cite{dieny1991giant,parkin1991giant}. Before the microfabrication process, the wafer was annealed in high vacuum at 280$^\circ$C for 1 h in a magnetic field of 5 kOe. The GMR strips with short and long axis of 2.5 and 70 $\mu$m were patterned using direct write laser lithography and ion beam milling. The wafer was patterned so that the easy axis of the magnetic free and reference layers were oriented along the shorter length of strips. In order to determine magnetizations, anisotropies and interlayer exchange coupling energies of the multilayer stack, we performed measurements using Vibrating Sample Magnetometer (VSM). From the magnetization hysteresis loop in high magnetic field we determined: saturation magnetization of the FL as 1.03 T, saturation magnetization of the RL as 1.65 T, exchange bias energy as 0.32 mJ/m$^2$. The uniaxial anisotropy energy of  0.8 kJ/m$^3$ was determined from the low
magnetic field hysteresis loop.

\subsection{Origin of the Oersted field}

\begin{figure}[!htbp]
\centering
\includegraphics[width=8.6cm]{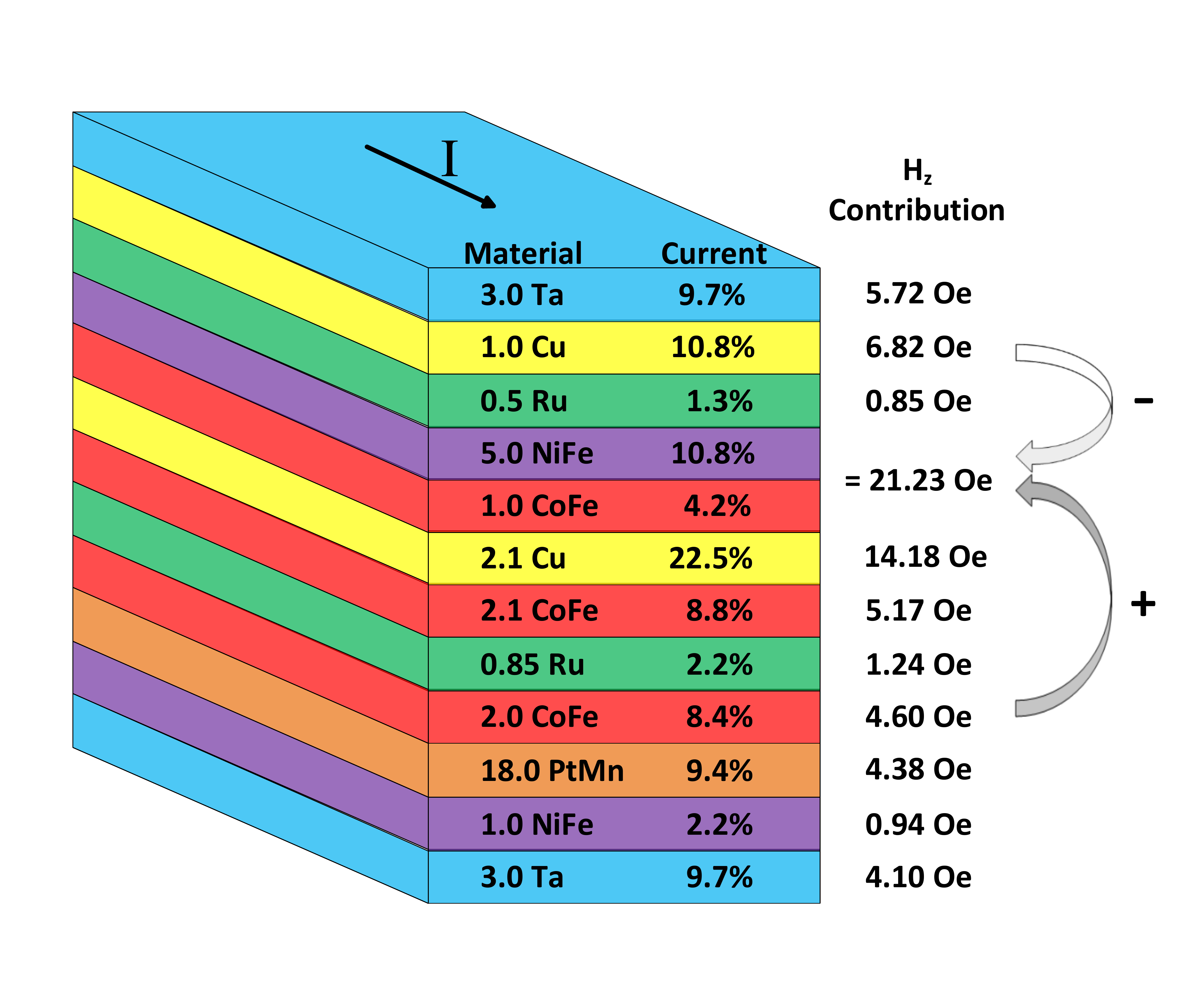}
\caption{Current distribution in GMR stack and magnetic field contribution from each layer. Note that the Oersted field contribution from the Cu spacer layer is dominant.}
\label{GMR_Strip}
\end{figure}

In order to investigate magnetization dynamics induced by the Oersted field we have performed micromagnetic simulations for a homogeneous ferromagnetic strip using Object Oriented Micromagnetic Framework.\cite{donahue1999oommf} We have found that if the current distribution in the single layer strip is uniform, the FMR modes excitated by alternating Oersted field do not contribute to the measured signal when averaged over the whole sample. In order to obtain a rectification signal, a non-compensated Oersted field component originating from some kind of symmetry breaking is needed. This conclusion stays in agreement with previous investigations,\cite{thiaville2008electrical} where mechanisms such as layer thickness effects described by Sondheimer-Fuchs model \cite{sondheimer1952mean} and contact regions influence were proposed to explain the origin of a nonzero resultant field. However, in the case of the SV-GMR strip we used, the asymmetry of the structure itself is sufficient to obtain a non-compensated Oersted field component. In order to calculate the magnitude of the resultant field, we have used the  estimations for thin film resistivities of the materials based on Refs.\onlinecite{desai1984electrical,eid2003enhancing,eid2002current,yuasa2002output,strelkov2003extension}.
 We have obtained resistance and current proportions for each layer, treating all layers as resistors connected in parallel, as depicted in Fig.\ref{GMR_Strip}. We have assumed that the current distribution is homogeneous within each layer. The total AC current amplitude has been taken from the experimental data. The resultant current distribution has been integrated over the whole sample using Biot-Savart law in order to obtain the Oersted field distribution. As seen in the Fig.\ref{GMR_Strip}, the total field in the free layer consists of several components originating from different layers, resulting in an uncompensated Oersted field with an amplitude of 21.23 Oe, which is sufficiently large to enable the FL magnetization coherent precession and the observed diode effect. Since it is only the resultant component of the Oersted field which induces the measured effect and the external field of 100 Oe generated by Helmholtz coils used in the experiment is strong enough to saturate the FL, we conclude that the macrospin approach is suitable for theoretical analysis.

\subsection{Experimental setup}

\begin{figure}[th]
    	       \centering
    	      \includegraphics[width=8.6cm]{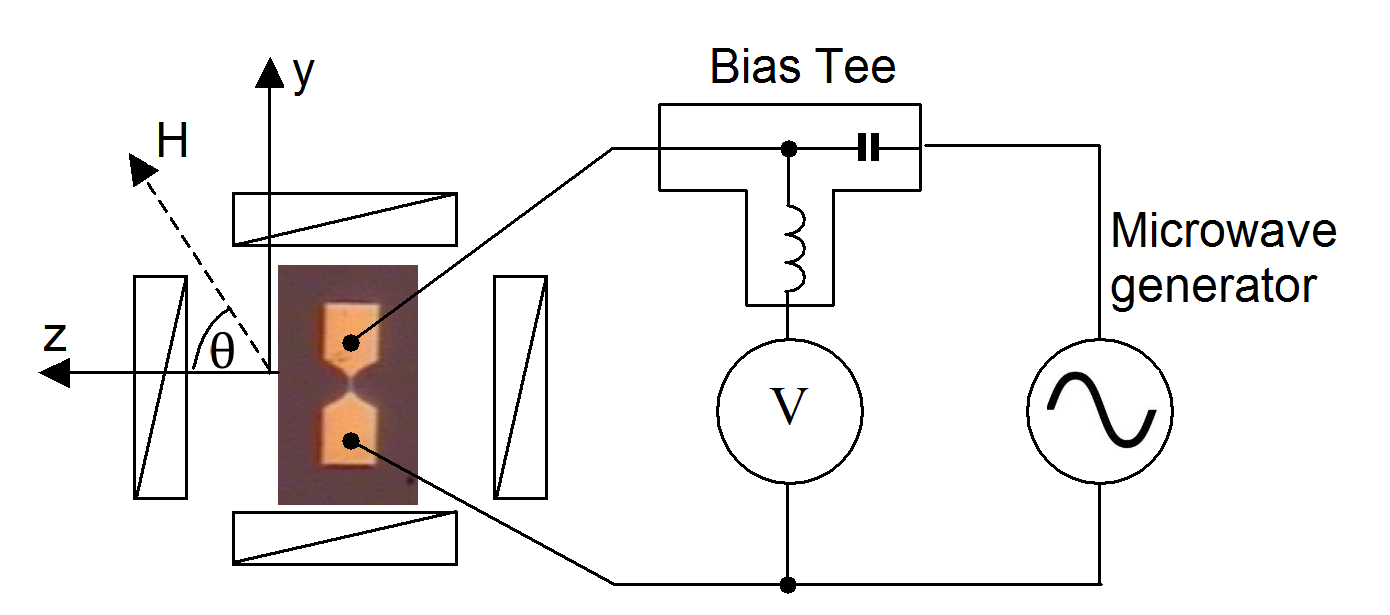}
    	       \caption{Schematic of the experimental setup for the spin diode effect detection. The axis $z$ denotes the exchange bias direction.}
    	       \label{schemat}
            \end{figure}

 The measurement setup consisted of a RF generator, a voltmeter, two pairs of Helmholtz coils oriented perpendicularly to each other, and a bias tee to separate the DC voltage component from the RF voltage (Fig.\ref{schemat}). We applied a microwave signal of 10 dBm from the generator to the strip via the ground-signal (GS) RF probe, and the DC voltage originating from the spin diode effect \cite{tulapurkar2005spin} was measured using magnetic field sweep along a given direction or the rotation of the magnetic field of constant magnitude within the $y-z$ plane. The angle $\theta$ between the magnetic field (or magnetization of the FL) and the z axis was varied by rotating magnetic field. The GMR measured in the strip was equal to 7.4$\%$.
Due to an impedance mismatch, the RF reflection coefficient was
$\gamma=\frac{R-Z_0}{R+Z_0}$ = 0.537, where $Z_0$ = 50 $\Omega$ is the impedance of the measurement system used. As a result, only $1-\gamma^2 = 71\%$ of the initial microwave power of 10 mW  (10 dBm) applied to the strip was actually absorbed, leading to the maximum current $I_{RF}$ = 6.5 mA.

\section{Results and discussion\label{sec:results}}

\subsection{Effective field}
In order to analyze the FMR dynamics and compare the experimental results with theoretical predictions, one needs to know the explicit form of the effective magnetic field $\vec{H}_{\rm eff}$ within the sample.
Based on experimental evidence, we can write the magnetic energy density in the form:
\begin{equation}
U =  K_\parallel \sin^2\theta - M_S (\vec{H}_{ext}\cdot\hat{e}_M -\frac{M_S}{2\mu_0} \hat{e}_M^T \mathrm{\hat{N}} \hat{e}_M+ H_{Oe}\hat{e}_z \cdot \hat{e}_M )
\label{eq:energia2}
\end{equation}
where $\hat{e}_M$ is a unit vector along the magnetization of the FL, $\mathrm{N}$ is demagnetization tensor in a standard form, $\rm K_\parallel$ describes the in-plane uniaxial magnetic anisotropy, and
$H_{Oe}$ is the amplitude of the uncompensated Oersted field in the $z$ direction. This form of the magnetostatic energy allows us to calculate derivatives in Eqs.(\ref{eq:resfreq}),(\ref{eq:sigma}) and (\ref{eq:vdcGMR}) at a given stationary angle $\theta_0$ set up by the external magnetic field. In particular, we are interested in the resonance frequency ($f_0=\omega_0/2\pi$) for $\theta=90^\circ$, at which the measured $V_{DC}$ signal is the strongest one:
\begin{eqnarray}
f_0 = \frac{\omega_0}{2\pi}= \frac{\gamma_e M_S }{2\pi} \left\{ \Gamma^{3/2} \left[ H_{ext} + \frac{M_S}{\mu_0}(N_x - N_y)\right]\right.  \nonumber \\
\left.\times \left( -H_{K\parallel} + (H_{ext} + \frac{M_S}{\mu_0}(N_z - N_y) \right)\right\}^{1/2}.
\label{eq:reson}
\end{eqnarray}
In the above equation, we expressed the uniaxial anisotropy in terms of the anisotropy field $H_{K\parallel} = 2 K_\parallel / M_S$.
Because of experimental conditions, we can neglect the symmetrical (antisymmetrical) contribution to $Z_1$ ($Z_2$), which is much smaller than its corresponding antisymmetrical (symmetrical) counterpart for $\psi = 0$,
and thus rewrite $Z_1$ ($Z_2$)  from Eq.(\ref{eq:vdcGMR}) for $\phi=\pi/2$ as:
\begin{widetext}
\begin{equation}
Z_1 \approx  M_S^2 \gamma_e \csc\theta \left( H_{ext}\sin^2\theta +\frac{M_S}{\mu_0} (N_x - N_y)\sin^2\theta \right)(\omega^2 - \omega^2 _0)
\end{equation}
\begin{equation}
Z_2 \approx  M_S^2 \gamma_e \csc\theta \left( H_{ext}\sin^2\theta + \frac{M_S}{\mu_0}(N_x - N_y)\sin^2\theta \right) \sigma\omega
\end{equation}
and then we express $V_{DC}$ signal as:
\begin{equation}
V_{DC} = \rm{A}\sin^2\theta \frac{1}{\Gamma((\omega^2 - \omega^2_0)^2 + \omega^2\sigma^2)}
 \left[ \cos\psi (\omega^2-\omega^2 _0) - \sin\psi \sigma\omega \right]
\label{eq:vdcGMRLAST}
\end{equation}
where:
\begin{equation}
\rm{A}\equiv \frac{\eta}{4} {M_S}^2 \Gamma^{1/2}\gamma^2 _e  I H_{Oe}\Delta R  \left( H_{ext}+\frac{M_S}{\mu_0}(N_x-N_y)\right)
\label{eq:amp}
\end{equation}
\end{widetext}

\subsection{ Magnetoresistance of SV-GMR strip}

\begin{figure}[!htbp]
    	       \centering
              \includegraphics[width=8.6cm]{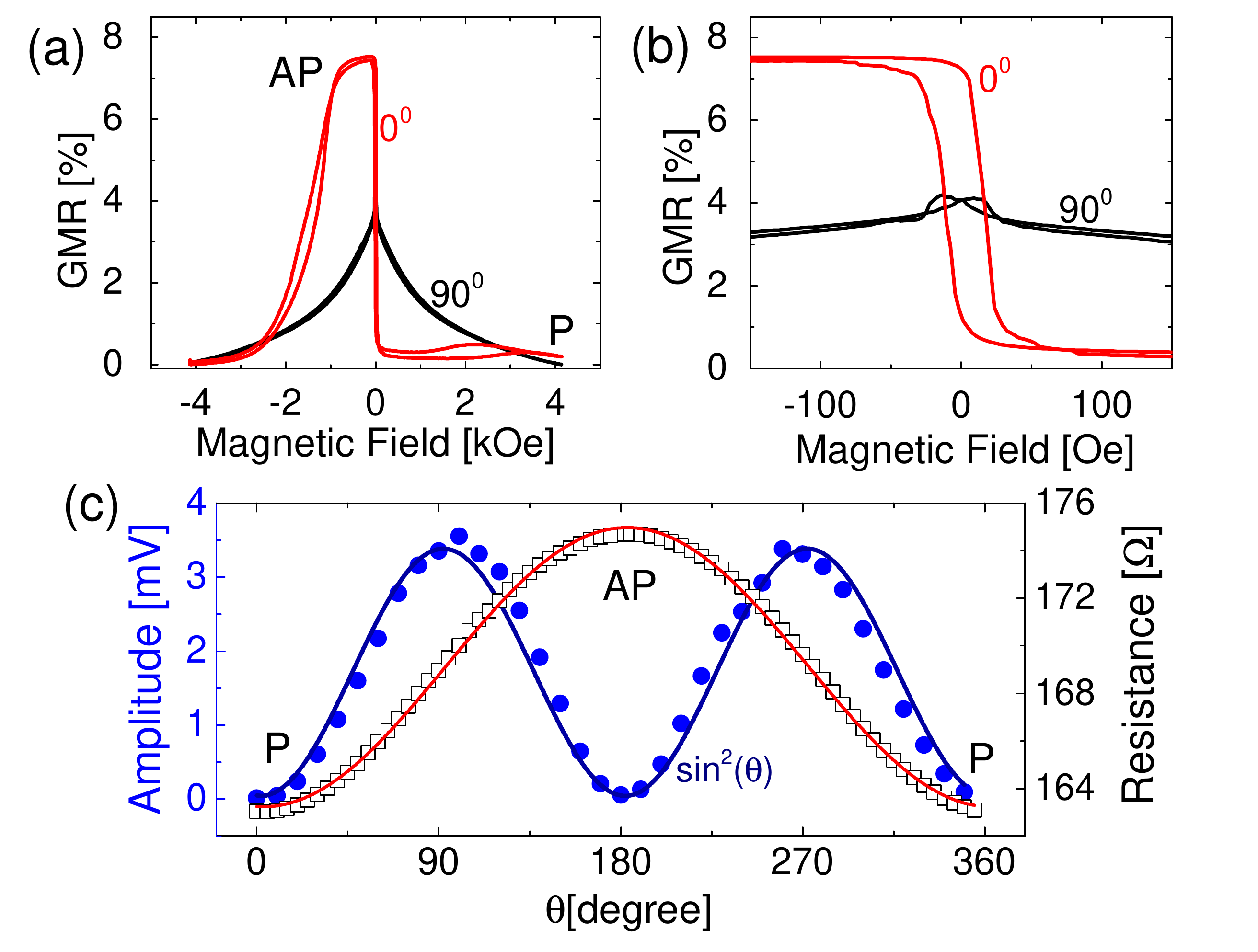}
    	       \caption{GMR loops measured in a wide magnetic field range (a) and  in a small magnetic field range (b). (c) Angular dependence of the resistance measured in 100 Oe (black squares), theoretical curve based on Eq.(\ref{RGMR}) (red solid line), spin diode DC voltage (blue points) and theoretical curve based on Eq.(\ref{eq:vdcGMRLAST}) (blue solid line).}
    	       \label{GMRLOOPS}
            \end{figure}

In our setup, the magnetoresistance has been measured in the -4 kOe to 4 kOe field range. Fig.\ref{GMRLOOPS}(a) shows relative changes of the resistance under sweeping field at $\theta$ = 0$^\circ$ (along the exchange bias direction) and  $\theta$ = 90$^\circ$ (perpendicular to the exchange bias direction), while in the Fig.\ref{GMRLOOPS}(b) the same loops are shown in the low field range. It should be emphasized that in our sample the AP state between FL and RL magnetizations has been fixed by the exchange bias field for values up to 1 kOe. The angular dependence of the magnetoresistance (Fig.\ref{GMRLOOPS}(c)) has been measured in rotating magnetic field of 100 Oe.

The amplitude for each angle has been calculated as a difference between maximum and minimum value of the spin diode DC voltage from the measured spectrum. It remains proportional to sin$^2$($\theta$), which is in accordance with Eq.(\ref{eq:vdcGMRLAST}).

\subsection{ Dynamics of SV-GMR strip}

\begin{figure}[!htbp]
    	       \centering
   	      \includegraphics[width=8.6cm]{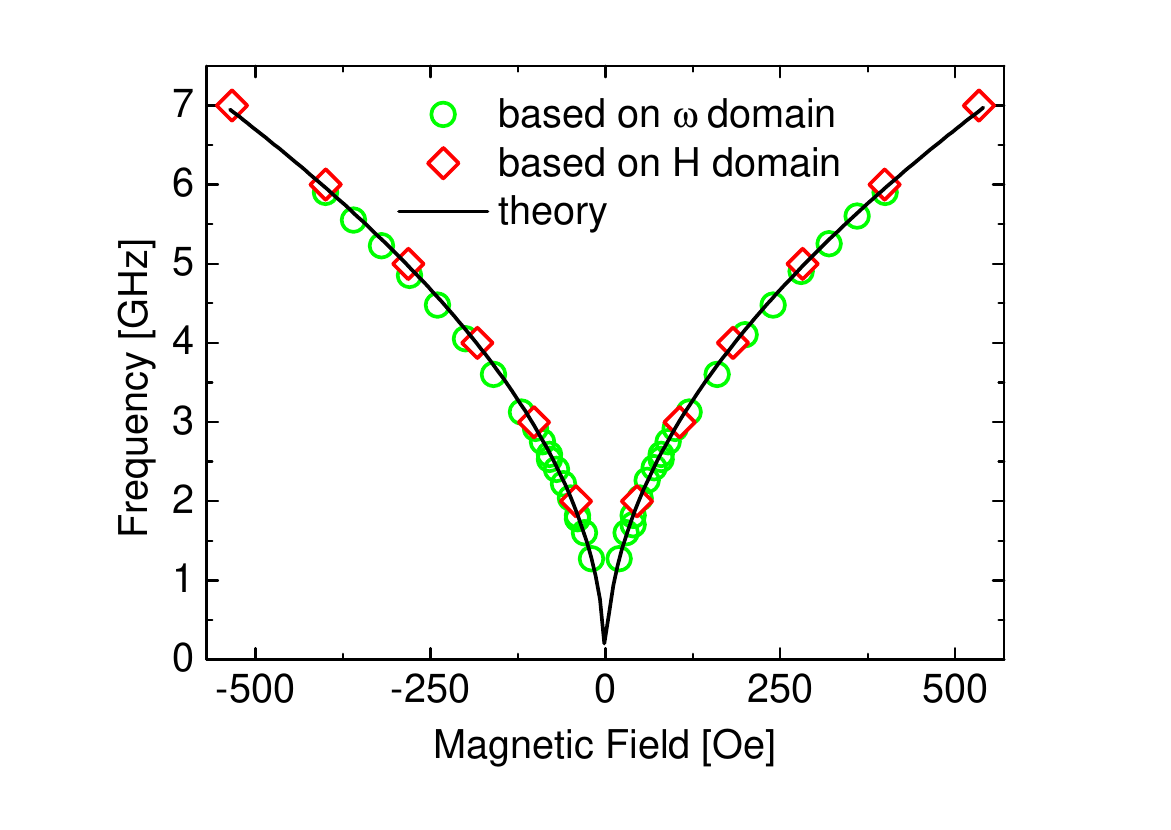}
    	       \caption{ Dispersion relation of the FL magnetization for the SV-GMR strip.}
    	       \label{kittel}
            \end{figure}

\begin{figure}[H]
    	       \centering
    	      \includegraphics[width=8.6cm]{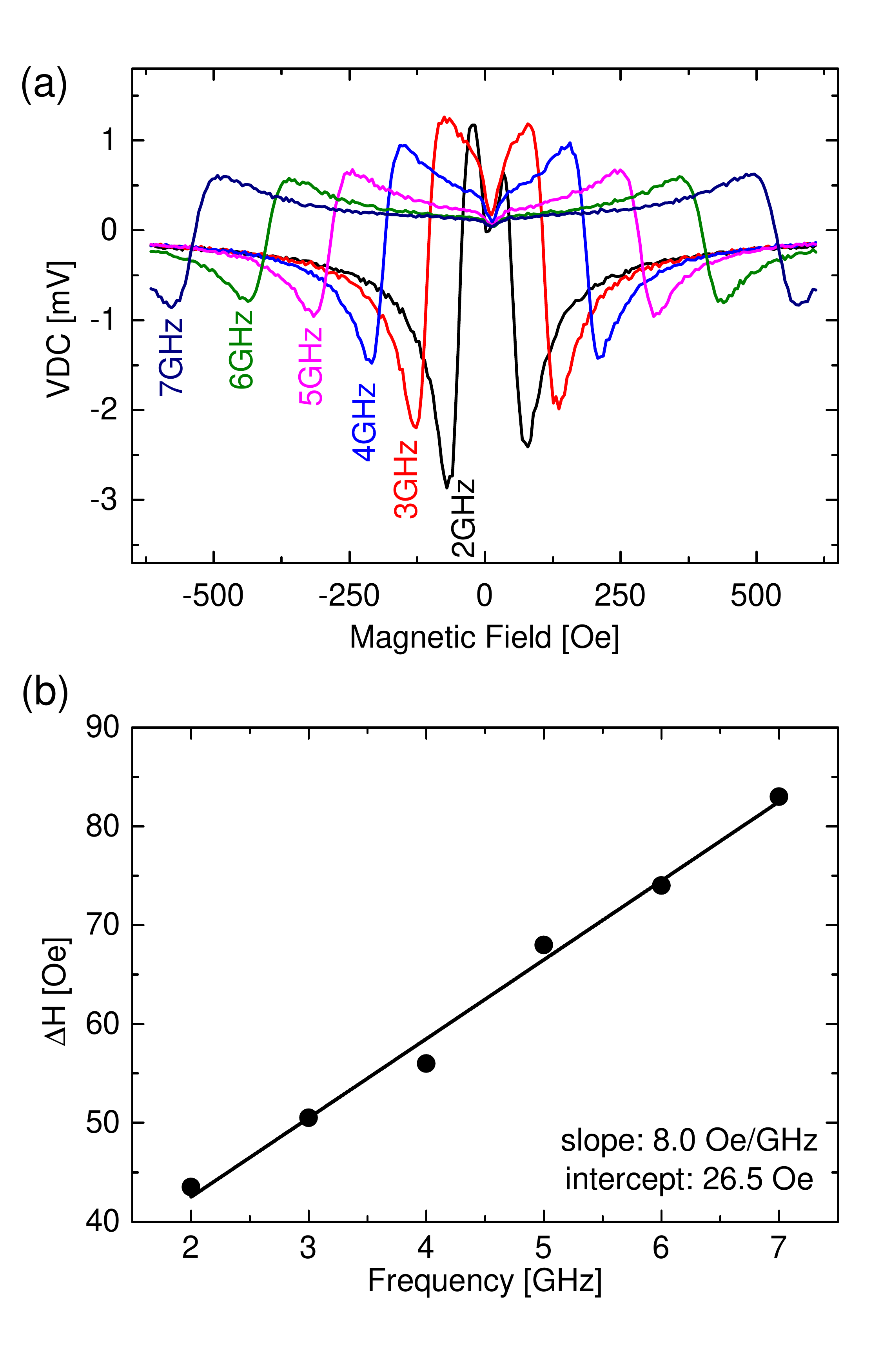}
    	       \caption{ (a) DC voltage from the spin diode effect as a function of magnetic field for various frequencies at $\theta$ = 90$^\circ$, (b) full width at half maximum ($\Delta H$) as a function of frequency.}
    	       \label{Hfield}
            \end{figure}

The FMR induced by the spin diode effect has been investigated as a function of frequency and magnetic field (in the range from -500 Oe to 500 Oe) as shown in Figs.\ref{kittel} and \ref{Hfield}(a).

This measurement was performed at $\theta$ = 90$^\circ$, where the amplitude of the FMR signal was maximal (Fig.\ref{GMRLOOPS}(c)). The resonance frequency shifts to higher values and the amplitude of the FMR signal decreases as the magnitude of the external magnetic field is increased (Fig.\ref{kittel},Fig.\ref{Hfield}). This behavior is in agreement with theoretical predictions (see Eq.(\ref{eq:reson}),(\ref{eq:vdcGMRLAST})). The theoretical dispersion relation  (Eq.\ref{eq:reson}) fits to the experimental
FMR data (Fig.\ref{kittel}) for values derived from VSM measurements (saturation magnetization of the FL 1.03 T, uniaxial anisotropy energy   0.8 kJ/m$^3$). The demagnetizing factors $N_x = 0.00187$, $N_y = 0.000065$ and $N_z = 0.998$ have been calculated with the use of analytical expressions for uniformly magnetized thin films, \cite{aharoni1998demagnetizing} taking into account the non-uniform (composite) character of the FL (1nm Co$_{90}$Fe$_{10}$ and 5nm Ni$_{81}$Fe$_{19}$).

 In order to determine the damping factor $\alpha$, the DC voltage signal has been measured as a function of field for several frequencies (Fig.\ref{Hfield}(a)). The linear dependence of $\Delta H$ as a function of frequency is shown in Fig.\ref{Hfield}(b). The damping factor has been determined by fitting a linear function to the measured data using the following equation:
\begin{equation}
\Delta H = \Delta H_0 + \alpha \frac{2 \pi f}{\gamma_e}.
\label{damping}
\end{equation}
Damping coefficient at $\theta$=90$^\circ$ calculated in this way is equal to 0.024.

\begin{figure}[!htbp]
    	       \centering
    	      \includegraphics[width=8.6cm]{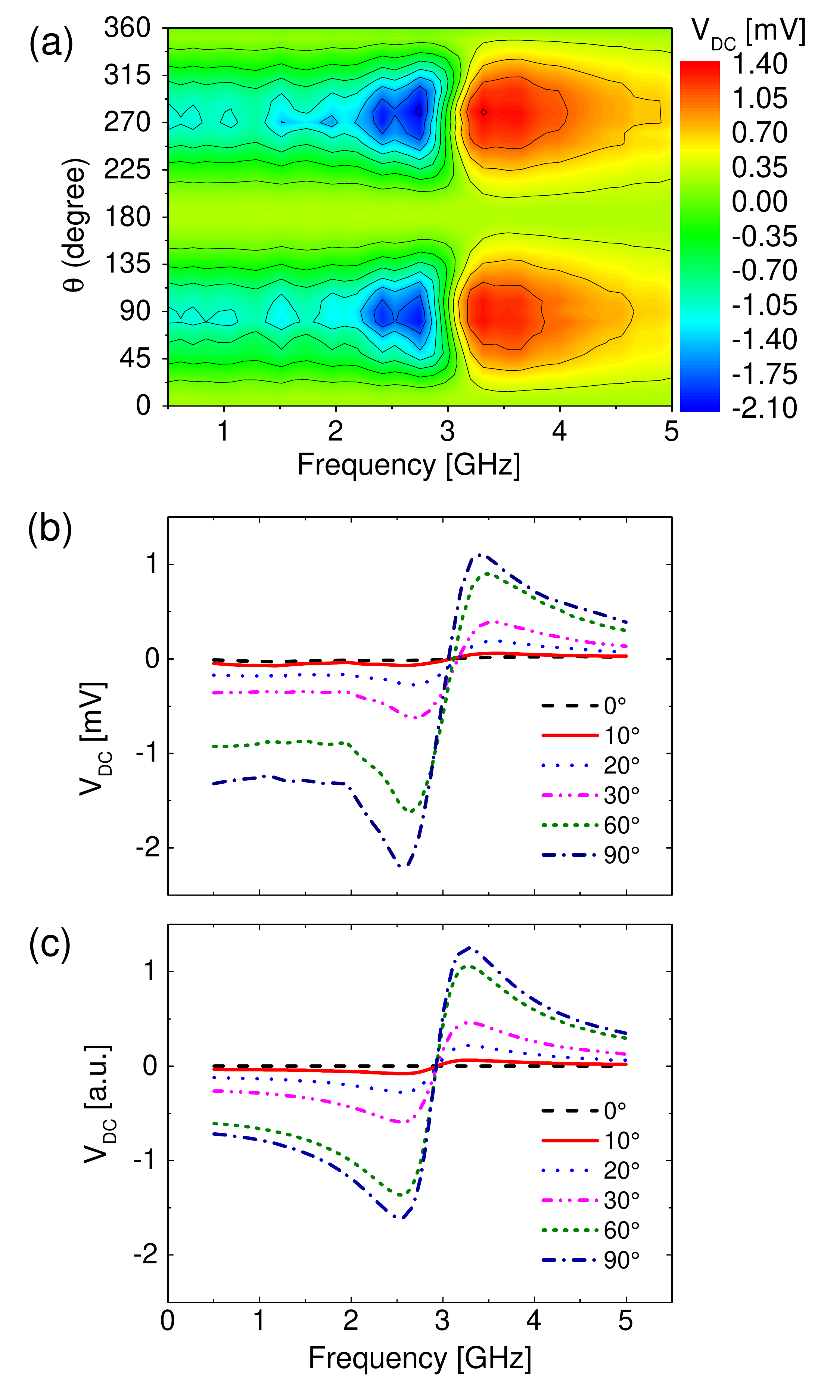}
    	       \caption{ (a) DC voltage originating from the spin diode effect as a function of magnetic field angle $\theta$ and (b) frequency. (c) Theoretical spectra predicted by Eq.(\ref{eq:vdcGMRLAST}). }
    	       \label{FMRanglemap}
\end{figure}

Finally, we have measured the $V_{DC}$ signal in the frequency domain. The spectra measured at different values of $\theta$ are presented in Fig.\ref{FMRanglemap}. Panel (a) shows frequency of FMR spectra in the full range of rotating  angles from 0$^\circ$ to 360$^\circ$. At $\theta$ equal to 0$^\circ$ and 180$^\circ$ the  FMR signal disappears, whereas at $\theta$ equal to 90$^\circ$ and 270$^\circ$ the signal achieves the maximum value.
The exact shapes of the experimental spectral lines are shown in Fig.\ref{FMRanglemap}(b). During the rotation of the magnetic field they retain their antisymmetrical character and their amplitudes follow a sine-squared dependence. Moreover, the resonance frequencies do not depend on the direction of the external field and the minima of the antisymmetrical curves have greater absolute values than their maxima.

In Fig.\ref{FMRanglemap}(c), we depict $V_{DC}$ curves plotted from Eq.(\ref{eq:vdcGMRLAST}) with the phase shift $\psi$ set to zero so that they are purely antisymmetrical, similarly to the experiment. One can see that all the features of experimental spectra mentioned above are reproduced. 

\subsection{Comparison of spin diode efficency in GMR and permalloy strips}

\begin{figure}[H]
    	       \centering
    	      \includegraphics[width=8.6cm]{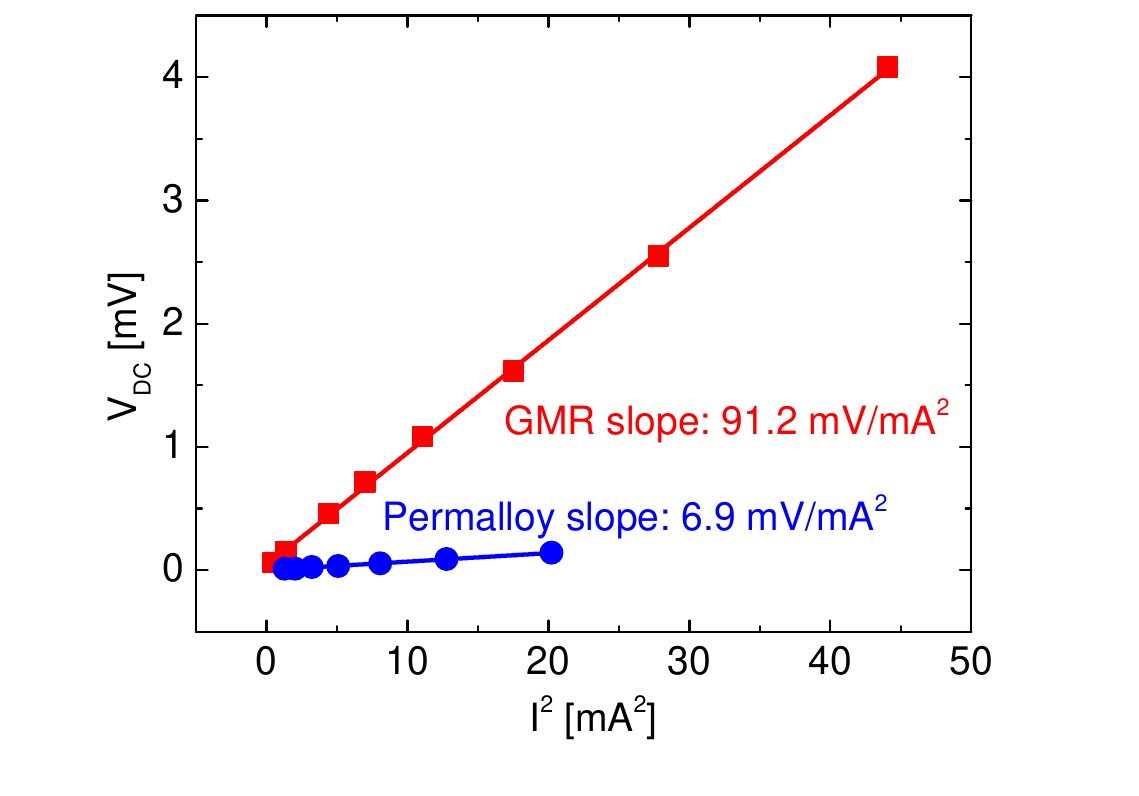}
    	       \caption{ Efficiency of the spin diode effect in SV-GMR and permalloy strips.}
    	       \label{power}
            \end{figure}

A similar experiment has also been conducted on permalloy strips. A 20 nm thick Ni$_{80}$Fe$_{20}$  layer has been deposited on an oxidized silicon wafer by magnetron sputtering.  Using electron-beam lithography and lift-off method, permalloy strips of dimensions similar to those of SV-GMR samples discussed above were fabricated. We found that, although the FMR signal still could be measured in this case, its amplitude was significantly smaller. The strip resistance was about two times larger (346 $\Omega$) and the magnetoresistance was an order of magnitude smaller than in the case of SV-GMR strip. The comparison between efficiencies of the spin diode effect for both types of sample shows that the efficiency in the SV-GMR strip is several times larger than in the permalloy one, as shown in Fig.\ref{power}.
This is consistent with previous reports on detection sensitivity being significantly larger in SV-GMR strips than in commonly used AMR devices.\cite{kleinlein2014using,yamaguchi2007rectification}
\\
                    
\section{Summary and conclusion\label{sec:summary}}
The spin diode effect was investigated both theoretically and experimentally in GMR strips. Measurements of both static magnetoresistance and the magnetization dynamics have been performed. In the GMR multilayer system, the symmetry of the current distribution is broken and a non-compensated Oersted field  appears in the FL, which enables for the $V_{DC}$ signal generation upon microwaves injection. The measured amplitude of the $V_{DC}$ signal has been shown to be significantly stronger than in commonly used AMR-based NiFe devices. We have provided a comprehensive theoretical model for calculations of the spin diode signal and used it to obtain $V_{DC}$ as function of frequency, external magnetic field, and angle at which the field is applied. The theoretical results are in a good quantitative agreement with the experimental data.
\section*{Acknowledgement}
We acknowledge Polish National Science Center grant Harmonia-DEC-2012/04/M/ST7/00799. Numerical calculations were supported in part by PL-GRID infrastructure. W.S. acknowledges the Foundation for Polish Science (FNP) scholarship under START Programme.


\bibliographystyle{apsrev4-1}

\bibliography{bibliography}

\begin{thebibliography}{43}%
\makeatletter
\providecommand \@ifxundefined [1]{%
 \@ifx{#1\undefined}
}%
\providecommand \@ifnum [1]{%
 \ifnum #1\expandafter \@firstoftwo
 \else \expandafter \@secondoftwo
 \fi
}%
\providecommand \@ifx [1]{%
 \ifx #1\expandafter \@firstoftwo
 \else \expandafter \@secondoftwo
 \fi
}%
\providecommand \natexlab [1]{#1}%
\providecommand \enquote  [1]{``#1''}%
\providecommand \bibnamefont  [1]{#1}%
\providecommand \bibfnamefont [1]{#1}%
\providecommand \citenamefont [1]{#1}%
\providecommand \href@noop [0]{\@secondoftwo}%
\providecommand \href [0]{\begingroup \@sanitize@url \@href}%
\providecommand \@href[1]{\@@startlink{#1}\@@href}%
\providecommand \@@href[1]{\endgroup#1\@@endlink}%
\providecommand \@sanitize@url [0]{\catcode `\\12\catcode `\$12\catcode
  `\&12\catcode `\#12\catcode `\^12\catcode `\_12\catcode `\%12\relax}%
\providecommand \@@startlink[1]{}%
\providecommand \@@endlink[0]{}%
\providecommand \url  [0]{\begingroup\@sanitize@url \@url }%
\providecommand \@url [1]{\endgroup\@href {#1}{\urlprefix }}%
\providecommand \urlprefix  [0]{URL }%
\providecommand \Eprint [0]{\href }%
\providecommand \doibase [0]{http://dx.doi.org/}%
\providecommand \selectlanguage [0]{\@gobble}%
\providecommand \bibinfo  [0]{\@secondoftwo}%
\providecommand \bibfield  [0]{\@secondoftwo}%
\providecommand \translation [1]{[#1]}%
\providecommand \BibitemOpen [0]{}%
\providecommand \bibitemStop [0]{}%
\providecommand \bibitemNoStop [0]{.\EOS\space}%
\providecommand \EOS [0]{\spacefactor3000\relax}%
\providecommand \BibitemShut  [1]{\csname bibitem#1\endcsname}%
\let\auto@bib@innerbib\@empty
\bibitem [{\citenamefont {Prokopenko}\ \emph {et~al.}(2013)\citenamefont
  {Prokopenko}, \citenamefont {Krivorotov}, \citenamefont {Meitzler},
  \citenamefont {Bankowski}, \citenamefont {Tiberkevich},\ and\ \citenamefont
  {Slavin}}]{prokopenko2013spin}%
  \BibitemOpen
  \bibfield  {author} {\bibinfo {author} {\bibfnamefont {O.}~\bibnamefont
  {Prokopenko}}, \bibinfo {author} {\bibfnamefont {I.}~\bibnamefont
  {Krivorotov}}, \bibinfo {author} {\bibfnamefont {T.}~\bibnamefont
  {Meitzler}}, \bibinfo {author} {\bibfnamefont {E.}~\bibnamefont {Bankowski}},
  \bibinfo {author} {\bibfnamefont {V.}~\bibnamefont {Tiberkevich}}, \ and\
  \bibinfo {author} {\bibfnamefont {A.}~\bibnamefont {Slavin}},\ }\href@noop {}
  {\bibfield  {journal} {\bibinfo  {journal} {Eds. S.O. Demokritov and A.N.
  Slavin, Magnonics: From Fundamentals to Applications. Series in Applied
  Physics}\ }\textbf {\bibinfo {volume} {125}},\ \bibinfo {pages} {143}
  (\bibinfo {year} {2013})}\BibitemShut {NoStop}%
\bibitem [{\citenamefont {Kim}(2012)}]{kima2012spin}%
  \BibitemOpen
  \bibfield  {author} {\bibinfo {author} {\bibfnamefont {J.-V.}\ \bibnamefont
  {Kim}},\ }\href@noop {} {\bibfield  {journal} {\bibinfo  {journal} {Solid
  State Phys.}\ }\textbf {\bibinfo {volume} {63}},\ \bibinfo {pages} {217}
  (\bibinfo {year} {2012})}\BibitemShut {NoStop}%
\bibitem [{\citenamefont {Sattler}(2010)}]{sattler2010handbook}%
  \BibitemOpen
  \bibfield  {author} {\bibinfo {author} {\bibfnamefont {K.}~\bibnamefont
  {Sattler}},\ }\href@noop {} {\emph {\bibinfo {title} {Handbook of
  nanophysics: Functional nanomaterials}}}\ (\bibinfo  {publisher} {CRC
  Press},\ \bibinfo {year} {2010})\BibitemShut {NoStop}%
\bibitem [{\citenamefont {Yamaguchi}\ \emph {et~al.}(2007)\citenamefont
  {Yamaguchi}, \citenamefont {Miyajima}, \citenamefont {Ono}, \citenamefont
  {Suzuki}, \citenamefont {Yuasa}, \citenamefont {Tulapurkar},\ and\
  \citenamefont {Nakatani}}]{yamaguchi2007rectification}%
  \BibitemOpen
  \bibfield  {author} {\bibinfo {author} {\bibfnamefont {A.}~\bibnamefont
  {Yamaguchi}}, \bibinfo {author} {\bibfnamefont {H.}~\bibnamefont {Miyajima}},
  \bibinfo {author} {\bibfnamefont {T.}~\bibnamefont {Ono}}, \bibinfo {author}
  {\bibfnamefont {Y.}~\bibnamefont {Suzuki}}, \bibinfo {author} {\bibfnamefont
  {S.}~\bibnamefont {Yuasa}}, \bibinfo {author} {\bibfnamefont
  {A.}~\bibnamefont {Tulapurkar}}, \ and\ \bibinfo {author} {\bibfnamefont
  {Y.}~\bibnamefont {Nakatani}},\ }\href@noop {} {\bibfield  {journal}
  {\bibinfo  {journal} {Appl. Phys. Lett.}\ }\textbf {\bibinfo {volume} {90}},\
  \bibinfo {pages} {182507} (\bibinfo {year} {2007})}\BibitemShut {NoStop}%
\bibitem [{\citenamefont {Harder}\ \emph {et~al.}(2011)\citenamefont {Harder},
  \citenamefont {Cao}, \citenamefont {Gui}, \citenamefont {Fan},\ and\
  \citenamefont {Hu}}]{harder2011analysis}%
  \BibitemOpen
  \bibfield  {author} {\bibinfo {author} {\bibfnamefont {M.}~\bibnamefont
  {Harder}}, \bibinfo {author} {\bibfnamefont {Z.}~\bibnamefont {Cao}},
  \bibinfo {author} {\bibfnamefont {Y.}~\bibnamefont {Gui}}, \bibinfo {author}
  {\bibfnamefont {X.}~\bibnamefont {Fan}}, \ and\ \bibinfo {author}
  {\bibfnamefont {C.-M.}\ \bibnamefont {Hu}},\ }\href@noop {} {\bibfield
  {journal} {\bibinfo  {journal} {Phys. Rev. B}\ }\textbf {\bibinfo {volume}
  {84}},\ \bibinfo {pages} {054423} (\bibinfo {year} {2011})}\BibitemShut
  {NoStop}%
\bibitem [{\citenamefont {Mecking}\ \emph {et~al.}(2007)\citenamefont
  {Mecking}, \citenamefont {Gui},\ and\ \citenamefont
  {Hu}}]{mecking2007microwave}%
  \BibitemOpen
  \bibfield  {author} {\bibinfo {author} {\bibfnamefont {N.}~\bibnamefont
  {Mecking}}, \bibinfo {author} {\bibfnamefont {Y.}~\bibnamefont {Gui}}, \ and\
  \bibinfo {author} {\bibfnamefont {C.-M.}\ \bibnamefont {Hu}},\ }\href@noop {}
  {\bibfield  {journal} {\bibinfo  {journal} {Phys. Rev. B}\ }\textbf {\bibinfo
  {volume} {76}},\ \bibinfo {pages} {224430} (\bibinfo {year}
  {2007})}\BibitemShut {NoStop}%
\bibitem [{\citenamefont {Kurebayashi}\ \emph {et~al.}(2014)\citenamefont
  {Kurebayashi}, \citenamefont {Sinova}, \citenamefont {Fang}, \citenamefont
  {Irvine}, \citenamefont {Skinner}, \citenamefont {Wunderlich}, \citenamefont
  {Nov{\'a}k}, \citenamefont {Campion}, \citenamefont {Gallagher},
  \citenamefont {Vehstedt} \emph {et~al.}}]{kurebayashi2014antidamping}%
  \BibitemOpen
  \bibfield  {author} {\bibinfo {author} {\bibfnamefont {H.}~\bibnamefont
  {Kurebayashi}}, \bibinfo {author} {\bibfnamefont {J.}~\bibnamefont {Sinova}},
  \bibinfo {author} {\bibfnamefont {D.}~\bibnamefont {Fang}}, \bibinfo {author}
  {\bibfnamefont {A.}~\bibnamefont {Irvine}}, \bibinfo {author} {\bibfnamefont
  {T.}~\bibnamefont {Skinner}}, \bibinfo {author} {\bibfnamefont
  {J.}~\bibnamefont {Wunderlich}}, \bibinfo {author} {\bibfnamefont
  {V.}~\bibnamefont {Nov{\'a}k}}, \bibinfo {author} {\bibfnamefont
  {R.}~\bibnamefont {Campion}}, \bibinfo {author} {\bibfnamefont
  {B.}~\bibnamefont {Gallagher}}, \bibinfo {author} {\bibfnamefont
  {E.}~\bibnamefont {Vehstedt}},  \emph {et~al.},\ }\href@noop {} {\bibfield
  {journal} {\bibinfo  {journal} {Nature Nanotech.}\ } (\bibinfo {year}
  {2014})}\BibitemShut {NoStop}%
\bibitem [{\citenamefont {Tulapurkar}\ \emph {et~al.}(2005)\citenamefont
  {Tulapurkar}, \citenamefont {Suzuki}, \citenamefont {Fukushima},
  \citenamefont {Kubota}, \citenamefont {Maehara}, \citenamefont {Tsunekawa},
  \citenamefont {Djayaprawira}, \citenamefont {Watanabe},\ and\ \citenamefont
  {Yuasa}}]{tulapurkar2005spin}%
  \BibitemOpen
  \bibfield  {author} {\bibinfo {author} {\bibfnamefont {A.}~\bibnamefont
  {Tulapurkar}}, \bibinfo {author} {\bibfnamefont {Y.}~\bibnamefont {Suzuki}},
  \bibinfo {author} {\bibfnamefont {A.}~\bibnamefont {Fukushima}}, \bibinfo
  {author} {\bibfnamefont {H.}~\bibnamefont {Kubota}}, \bibinfo {author}
  {\bibfnamefont {H.}~\bibnamefont {Maehara}}, \bibinfo {author} {\bibfnamefont
  {K.}~\bibnamefont {Tsunekawa}}, \bibinfo {author} {\bibfnamefont
  {D.}~\bibnamefont {Djayaprawira}}, \bibinfo {author} {\bibfnamefont
  {N.}~\bibnamefont {Watanabe}}, \ and\ \bibinfo {author} {\bibfnamefont
  {S.}~\bibnamefont {Yuasa}},\ }\href@noop {} {\bibfield  {journal} {\bibinfo
  {journal} {Nature}\ }\textbf {\bibinfo {volume} {438}},\ \bibinfo {pages}
  {339} (\bibinfo {year} {2005})}\BibitemShut {NoStop}%
\bibitem [{\citenamefont {Sankey}\ \emph {et~al.}(2006)\citenamefont {Sankey},
  \citenamefont {Braganca}, \citenamefont {Garcia}, \citenamefont {Krivorotov},
  \citenamefont {Buhrman},\ and\ \citenamefont {Ralph}}]{sankey2006spin}%
  \BibitemOpen
  \bibfield  {author} {\bibinfo {author} {\bibfnamefont {J.}~\bibnamefont
  {Sankey}}, \bibinfo {author} {\bibfnamefont {P.}~\bibnamefont {Braganca}},
  \bibinfo {author} {\bibfnamefont {A.}~\bibnamefont {Garcia}}, \bibinfo
  {author} {\bibfnamefont {I.}~\bibnamefont {Krivorotov}}, \bibinfo {author}
  {\bibfnamefont {R.}~\bibnamefont {Buhrman}}, \ and\ \bibinfo {author}
  {\bibfnamefont {D.}~\bibnamefont {Ralph}},\ }\href@noop {} {\bibfield
  {journal} {\bibinfo  {journal} {Phys. Rev. Lett.}\ }\textbf {\bibinfo
  {volume} {96}},\ \bibinfo {pages} {227601} (\bibinfo {year}
  {2006})}\BibitemShut {NoStop}%
\bibitem [{\citenamefont {Sankey}\ \emph {et~al.}(2007)\citenamefont {Sankey},
  \citenamefont {Cui}, \citenamefont {Sun}, \citenamefont {Slonczewski},
  \citenamefont {Buhrman},\ and\ \citenamefont
  {Ralph}}]{sankey2007measurement}%
  \BibitemOpen
  \bibfield  {author} {\bibinfo {author} {\bibfnamefont {J.}~\bibnamefont
  {Sankey}}, \bibinfo {author} {\bibfnamefont {Y.-T.}\ \bibnamefont {Cui}},
  \bibinfo {author} {\bibfnamefont {J.}~\bibnamefont {Sun}}, \bibinfo {author}
  {\bibfnamefont {J.}~\bibnamefont {Slonczewski}}, \bibinfo {author}
  {\bibfnamefont {R.}~\bibnamefont {Buhrman}}, \ and\ \bibinfo {author}
  {\bibfnamefont {D.}~\bibnamefont {Ralph}},\ }\href@noop {} {\bibfield
  {journal} {\bibinfo  {journal} {Nat. Phys.}\ }\textbf {\bibinfo {volume}
  {4}},\ \bibinfo {pages} {67} (\bibinfo {year} {2007})}\BibitemShut {NoStop}%
\bibitem [{\citenamefont {Skowro{\'n}ski}\ \emph {et~al.}(2013)\citenamefont
  {Skowro{\'n}ski}, \citenamefont {Czapkiewicz}, \citenamefont {Frankowski},
  \citenamefont {Wrona}, \citenamefont {Stobiecki}, \citenamefont {Reiss},
  \citenamefont {Chalapat}, \citenamefont {Paraoanu},\ and\ \citenamefont {van
  Dijken}}]{skowronski2013influence}%
  \BibitemOpen
  \bibfield  {author} {\bibinfo {author} {\bibfnamefont {W.}~\bibnamefont
  {Skowro{\'n}ski}}, \bibinfo {author} {\bibfnamefont {M.}~\bibnamefont
  {Czapkiewicz}}, \bibinfo {author} {\bibfnamefont {M.}~\bibnamefont
  {Frankowski}}, \bibinfo {author} {\bibfnamefont {J.}~\bibnamefont {Wrona}},
  \bibinfo {author} {\bibfnamefont {T.}~\bibnamefont {Stobiecki}}, \bibinfo
  {author} {\bibfnamefont {G.}~\bibnamefont {Reiss}}, \bibinfo {author}
  {\bibfnamefont {K.}~\bibnamefont {Chalapat}}, \bibinfo {author}
  {\bibfnamefont {G.}~\bibnamefont {Paraoanu}}, \ and\ \bibinfo {author}
  {\bibfnamefont {S.}~\bibnamefont {van Dijken}},\ }\href@noop {} {\bibfield
  {journal} {\bibinfo  {journal} {Phys. Rev. B}\ }\textbf {\bibinfo {volume}
  {87}},\ \bibinfo {pages} {094419} (\bibinfo {year} {2013})}\BibitemShut
  {NoStop}%
\bibitem [{\citenamefont {Skowro{\'n}ski}\ \emph {et~al.}(2014)\citenamefont
  {Skowro{\'n}ski}, \citenamefont {Frankowski}, \citenamefont {Wrona},
  \citenamefont {Stobiecki}, \citenamefont {Ogrodnik},\ and\ \citenamefont
  {Barna{\'s}}}]{skowronski2014spin}%
  \BibitemOpen
  \bibfield  {author} {\bibinfo {author} {\bibfnamefont {W.}~\bibnamefont
  {Skowro{\'n}ski}}, \bibinfo {author} {\bibfnamefont {M.}~\bibnamefont
  {Frankowski}}, \bibinfo {author} {\bibfnamefont {J.}~\bibnamefont {Wrona}},
  \bibinfo {author} {\bibfnamefont {T.}~\bibnamefont {Stobiecki}}, \bibinfo
  {author} {\bibfnamefont {P.}~\bibnamefont {Ogrodnik}}, \ and\ \bibinfo
  {author} {\bibfnamefont {J.}~\bibnamefont {Barna{\'s}}},\ }\href@noop {}
  {\bibfield  {journal} {\bibinfo  {journal} {Appl. Phys. Lett.}\ }\textbf
  {\bibinfo {volume} {105}},\ \bibinfo {pages} {072409} (\bibinfo {year}
  {2014})}\BibitemShut {NoStop}%
\bibitem [{\citenamefont {Kubota}\ \emph {et~al.}(2007)\citenamefont {Kubota},
  \citenamefont {Fukushima}, \citenamefont {Yakushiji}, \citenamefont
  {Nagahama}, \citenamefont {Yuasa}, \citenamefont {Ando}, \citenamefont
  {Maehara}, \citenamefont {Nagamine}, \citenamefont {Tsunekawa}, \citenamefont
  {Djayaprawira} \emph {et~al.}}]{kubota2007quantitative}%
  \BibitemOpen
  \bibfield  {author} {\bibinfo {author} {\bibfnamefont {H.}~\bibnamefont
  {Kubota}}, \bibinfo {author} {\bibfnamefont {A.}~\bibnamefont {Fukushima}},
  \bibinfo {author} {\bibfnamefont {K.}~\bibnamefont {Yakushiji}}, \bibinfo
  {author} {\bibfnamefont {T.}~\bibnamefont {Nagahama}}, \bibinfo {author}
  {\bibfnamefont {S.}~\bibnamefont {Yuasa}}, \bibinfo {author} {\bibfnamefont
  {K.}~\bibnamefont {Ando}}, \bibinfo {author} {\bibfnamefont {H.}~\bibnamefont
  {Maehara}}, \bibinfo {author} {\bibfnamefont {Y.}~\bibnamefont {Nagamine}},
  \bibinfo {author} {\bibfnamefont {K.}~\bibnamefont {Tsunekawa}}, \bibinfo
  {author} {\bibfnamefont {D.}~\bibnamefont {Djayaprawira}},  \emph {et~al.},\
  }\href@noop {} {\bibfield  {journal} {\bibinfo  {journal} {Nature Phys.}\
  }\textbf {\bibinfo {volume} {4}},\ \bibinfo {pages} {37} (\bibinfo {year}
  {2007})}\BibitemShut {NoStop}%
\bibitem [{\citenamefont {Gui}\ \emph {et~al.}(2005)\citenamefont {Gui},
  \citenamefont {Holland}, \citenamefont {Mecking},\ and\ \citenamefont
  {Hu}}]{gui2005resonances}%
  \BibitemOpen
  \bibfield  {author} {\bibinfo {author} {\bibfnamefont {Y.}~\bibnamefont
  {Gui}}, \bibinfo {author} {\bibfnamefont {S.}~\bibnamefont {Holland}},
  \bibinfo {author} {\bibfnamefont {N.}~\bibnamefont {Mecking}}, \ and\
  \bibinfo {author} {\bibfnamefont {C.-M.}\ \bibnamefont {Hu}},\ }\href@noop {}
  {\bibfield  {journal} {\bibinfo  {journal} {Phys. Rev. Lett.}\ }\textbf
  {\bibinfo {volume} {95}},\ \bibinfo {pages} {056807} (\bibinfo {year}
  {2005})}\BibitemShut {NoStop}%
\bibitem [{\citenamefont {Costache}\ \emph
  {et~al.}(2006{\natexlab{a}})\citenamefont {Costache}, \citenamefont {Watts},
  \citenamefont {Sladkov}, \citenamefont {Van~der Wal},\ and\ \citenamefont
  {Van~Wees}}]{costache2006large}%
  \BibitemOpen
  \bibfield  {author} {\bibinfo {author} {\bibfnamefont {M.}~\bibnamefont
  {Costache}}, \bibinfo {author} {\bibfnamefont {S.}~\bibnamefont {Watts}},
  \bibinfo {author} {\bibfnamefont {M.}~\bibnamefont {Sladkov}}, \bibinfo
  {author} {\bibfnamefont {C.}~\bibnamefont {Van~der Wal}}, \ and\ \bibinfo
  {author} {\bibfnamefont {B.}~\bibnamefont {Van~Wees}},\ }\href@noop {}
  {\bibfield  {journal} {\bibinfo  {journal} {Appl. Phys. Lett.}\ }\textbf
  {\bibinfo {volume} {89}},\ \bibinfo {pages} {232115} (\bibinfo {year}
  {2006}{\natexlab{a}})}\BibitemShut {NoStop}%
\bibitem [{\citenamefont {Gui}\ \emph {et~al.}(2007)\citenamefont {Gui},
  \citenamefont {Mecking}, \citenamefont {Zhou}, \citenamefont {Williams},\
  and\ \citenamefont {Hu}}]{gui2007realization}%
  \BibitemOpen
  \bibfield  {author} {\bibinfo {author} {\bibfnamefont {Y.}~\bibnamefont
  {Gui}}, \bibinfo {author} {\bibfnamefont {N.}~\bibnamefont {Mecking}},
  \bibinfo {author} {\bibfnamefont {X.}~\bibnamefont {Zhou}}, \bibinfo {author}
  {\bibfnamefont {G.}~\bibnamefont {Williams}}, \ and\ \bibinfo {author}
  {\bibfnamefont {C.-M.}\ \bibnamefont {Hu}},\ }\href@noop {} {\bibfield
  {journal} {\bibinfo  {journal} {Phys. Rev. Lett.}\ }\textbf {\bibinfo
  {volume} {98}},\ \bibinfo {pages} {107602} (\bibinfo {year}
  {2007})}\BibitemShut {NoStop}%
\bibitem [{\citenamefont {Hui}\ \emph {et~al.}(2008)\citenamefont {Hui},
  \citenamefont {Wirthmann}, \citenamefont {Gui}, \citenamefont {Tian},
  \citenamefont {Jin}, \citenamefont {Chen}, \citenamefont {Shen},\ and\
  \citenamefont {Hu}}]{hui2008electric}%
  \BibitemOpen
  \bibfield  {author} {\bibinfo {author} {\bibfnamefont {X.}~\bibnamefont
  {Hui}}, \bibinfo {author} {\bibfnamefont {A.}~\bibnamefont {Wirthmann}},
  \bibinfo {author} {\bibfnamefont {Y.}~\bibnamefont {Gui}}, \bibinfo {author}
  {\bibfnamefont {Y.}~\bibnamefont {Tian}}, \bibinfo {author} {\bibfnamefont
  {X.}~\bibnamefont {Jin}}, \bibinfo {author} {\bibfnamefont {Z.}~\bibnamefont
  {Chen}}, \bibinfo {author} {\bibfnamefont {S.}~\bibnamefont {Shen}}, \ and\
  \bibinfo {author} {\bibfnamefont {C.-M.}\ \bibnamefont {Hu}},\ }\href@noop {}
  {\bibfield  {journal} {\bibinfo  {journal} {Appl. Phys. Lett.}\ }\textbf
  {\bibinfo {volume} {93}},\ \bibinfo {pages} {232502} (\bibinfo {year}
  {2008})}\BibitemShut {NoStop}%
\bibitem [{\citenamefont {Goennenwein}\ \emph {et~al.}(2007)\citenamefont
  {Goennenwein}, \citenamefont {Schink}, \citenamefont {Brandlmaier},
  \citenamefont {Boger}, \citenamefont {Opel}, \citenamefont {Gross},
  \citenamefont {Keizer}, \citenamefont {Klapwijk}, \citenamefont {Gupta},
  \citenamefont {Huebl} \emph {et~al.}}]{goennenwein2007electrically}%
  \BibitemOpen
  \bibfield  {author} {\bibinfo {author} {\bibfnamefont {S.}~\bibnamefont
  {Goennenwein}}, \bibinfo {author} {\bibfnamefont {S.}~\bibnamefont {Schink}},
  \bibinfo {author} {\bibfnamefont {A.}~\bibnamefont {Brandlmaier}}, \bibinfo
  {author} {\bibfnamefont {A.}~\bibnamefont {Boger}}, \bibinfo {author}
  {\bibfnamefont {M.}~\bibnamefont {Opel}}, \bibinfo {author} {\bibfnamefont
  {R.}~\bibnamefont {Gross}}, \bibinfo {author} {\bibfnamefont
  {R.}~\bibnamefont {Keizer}}, \bibinfo {author} {\bibfnamefont
  {T.}~\bibnamefont {Klapwijk}}, \bibinfo {author} {\bibfnamefont
  {A.}~\bibnamefont {Gupta}}, \bibinfo {author} {\bibfnamefont
  {H.}~\bibnamefont {Huebl}},  \emph {et~al.},\ }\href@noop {} {\bibfield
  {journal} {\bibinfo  {journal} {Appl. Phys. Lett.}\ }\textbf {\bibinfo
  {volume} {90}},\ \bibinfo {pages} {162507} (\bibinfo {year}
  {2007})}\BibitemShut {NoStop}%
\bibitem [{\citenamefont {Wirthmann}\ \emph {et~al.}(2008)\citenamefont
  {Wirthmann}, \citenamefont {Hui}, \citenamefont {Mecking}, \citenamefont
  {Gui}, \citenamefont {Chakraborty}, \citenamefont {Hu}, \citenamefont
  {Reinwald}, \citenamefont {Sch{\"u}ller},\ and\ \citenamefont
  {Wegscheider}}]{wirthmann2008broadband}%
  \BibitemOpen
  \bibfield  {author} {\bibinfo {author} {\bibfnamefont {A.}~\bibnamefont
  {Wirthmann}}, \bibinfo {author} {\bibfnamefont {X.}~\bibnamefont {Hui}},
  \bibinfo {author} {\bibfnamefont {N.}~\bibnamefont {Mecking}}, \bibinfo
  {author} {\bibfnamefont {Y.}~\bibnamefont {Gui}}, \bibinfo {author}
  {\bibfnamefont {T.}~\bibnamefont {Chakraborty}}, \bibinfo {author}
  {\bibfnamefont {C.-M.}\ \bibnamefont {Hu}}, \bibinfo {author} {\bibfnamefont
  {M.}~\bibnamefont {Reinwald}}, \bibinfo {author} {\bibfnamefont
  {C.}~\bibnamefont {Sch{\"u}ller}}, \ and\ \bibinfo {author} {\bibfnamefont
  {W.}~\bibnamefont {Wegscheider}},\ }\href@noop {} {\bibfield  {journal}
  {\bibinfo  {journal} {Appl. Phys. Lett.}\ }\textbf {\bibinfo {volume} {92}},\
  \bibinfo {pages} {232106} (\bibinfo {year} {2008})}\BibitemShut {NoStop}%
\bibitem [{\citenamefont {Costache}\ \emph
  {et~al.}(2006{\natexlab{b}})\citenamefont {Costache}, \citenamefont
  {Sladkov}, \citenamefont {Watts}, \citenamefont {van~der Wal},\ and\
  \citenamefont {van Wees}}]{costache2006electrical}%
  \BibitemOpen
  \bibfield  {author} {\bibinfo {author} {\bibfnamefont {M.}~\bibnamefont
  {Costache}}, \bibinfo {author} {\bibfnamefont {M.}~\bibnamefont {Sladkov}},
  \bibinfo {author} {\bibfnamefont {S.}~\bibnamefont {Watts}}, \bibinfo
  {author} {\bibfnamefont {C.}~\bibnamefont {van~der Wal}}, \ and\ \bibinfo
  {author} {\bibfnamefont {B.}~\bibnamefont {van Wees}},\ }\href@noop {}
  {\bibfield  {journal} {\bibinfo  {journal} {Phys. Rev. Lett.}\ }\textbf
  {\bibinfo {volume} {97}},\ \bibinfo {pages} {216603} (\bibinfo {year}
  {2006}{\natexlab{b}})}\BibitemShut {NoStop}%
\bibitem [{\citenamefont {Saitoh}\ \emph {et~al.}(2006)\citenamefont {Saitoh},
  \citenamefont {Ueda}, \citenamefont {Miyajima},\ and\ \citenamefont
  {Tatara}}]{saitoh2006conversion}%
  \BibitemOpen
  \bibfield  {author} {\bibinfo {author} {\bibfnamefont {E.}~\bibnamefont
  {Saitoh}}, \bibinfo {author} {\bibfnamefont {M.}~\bibnamefont {Ueda}},
  \bibinfo {author} {\bibfnamefont {H.}~\bibnamefont {Miyajima}}, \ and\
  \bibinfo {author} {\bibfnamefont {G.}~\bibnamefont {Tatara}},\ }\href@noop {}
  {\bibfield  {journal} {\bibinfo  {journal} {Appl. Phys. Lett.}\ }\textbf
  {\bibinfo {volume} {88}},\ \bibinfo {pages} {182509} (\bibinfo {year}
  {2006})}\BibitemShut {NoStop}%
\bibitem [{\citenamefont {Mosendz}\ \emph
  {et~al.}(2010{\natexlab{a}})\citenamefont {Mosendz}, \citenamefont {Pearson},
  \citenamefont {Fradin}, \citenamefont {Bauer}, \citenamefont {Bader},\ and\
  \citenamefont {Hoffmann}}]{mosendz2010quantifying}%
  \BibitemOpen
  \bibfield  {author} {\bibinfo {author} {\bibfnamefont {O.}~\bibnamefont
  {Mosendz}}, \bibinfo {author} {\bibfnamefont {J.}~\bibnamefont {Pearson}},
  \bibinfo {author} {\bibfnamefont {F.}~\bibnamefont {Fradin}}, \bibinfo
  {author} {\bibfnamefont {G.}~\bibnamefont {Bauer}}, \bibinfo {author}
  {\bibfnamefont {S.}~\bibnamefont {Bader}}, \ and\ \bibinfo {author}
  {\bibfnamefont {A.}~\bibnamefont {Hoffmann}},\ }\href@noop {} {\bibfield
  {journal} {\bibinfo  {journal} {Phys. Rev. Lett.}\ }\textbf {\bibinfo
  {volume} {104}},\ \bibinfo {pages} {046601} (\bibinfo {year}
  {2010}{\natexlab{a}})}\BibitemShut {NoStop}%
\bibitem [{\citenamefont {Mosendz}\ \emph
  {et~al.}(2010{\natexlab{b}})\citenamefont {Mosendz}, \citenamefont
  {Vlaminck}, \citenamefont {Pearson}, \citenamefont {Fradin}, \citenamefont
  {Bauer}, \citenamefont {Bader},\ and\ \citenamefont
  {Hoffmann}}]{mosendz2010detection}%
  \BibitemOpen
  \bibfield  {author} {\bibinfo {author} {\bibfnamefont {O.}~\bibnamefont
  {Mosendz}}, \bibinfo {author} {\bibfnamefont {V.}~\bibnamefont {Vlaminck}},
  \bibinfo {author} {\bibfnamefont {J.}~\bibnamefont {Pearson}}, \bibinfo
  {author} {\bibfnamefont {F.}~\bibnamefont {Fradin}}, \bibinfo {author}
  {\bibfnamefont {G.}~\bibnamefont {Bauer}}, \bibinfo {author} {\bibfnamefont
  {S.}~\bibnamefont {Bader}}, \ and\ \bibinfo {author} {\bibfnamefont
  {A.}~\bibnamefont {Hoffmann}},\ }\href@noop {} {\bibfield  {journal}
  {\bibinfo  {journal} {Phys. Rev. B}\ }\textbf {\bibinfo {volume} {82}},\
  \bibinfo {pages} {214403} (\bibinfo {year} {2010}{\natexlab{b}})}\BibitemShut
  {NoStop}%
\bibitem [{\citenamefont {Saraiva}\ \emph {et~al.}(2010)\citenamefont
  {Saraiva}, \citenamefont {Nogaret}, \citenamefont {Portal}, \citenamefont
  {Beere},\ and\ \citenamefont {Ritchie}}]{saraiva2010dipolar}%
  \BibitemOpen
  \bibfield  {author} {\bibinfo {author} {\bibfnamefont {P.}~\bibnamefont
  {Saraiva}}, \bibinfo {author} {\bibfnamefont {A.}~\bibnamefont {Nogaret}},
  \bibinfo {author} {\bibfnamefont {J.}~\bibnamefont {Portal}}, \bibinfo
  {author} {\bibfnamefont {H.}~\bibnamefont {Beere}}, \ and\ \bibinfo {author}
  {\bibfnamefont {D.}~\bibnamefont {Ritchie}},\ }\href@noop {} {\bibfield
  {journal} {\bibinfo  {journal} {Phys. Rev. B}\ }\textbf {\bibinfo {volume}
  {82}},\ \bibinfo {pages} {224417} (\bibinfo {year} {2010})}\BibitemShut
  {NoStop}%
\bibitem [{\citenamefont {Kajiwara}\ \emph {et~al.}(2010)\citenamefont
  {Kajiwara}, \citenamefont {Harii}, \citenamefont {Takahashi}, \citenamefont
  {Ohe}, \citenamefont {Uchida}, \citenamefont {Mizuguchi}, \citenamefont
  {Umezawa}, \citenamefont {Kawai}, \citenamefont {Ando}, \citenamefont
  {Takanashi} \emph {et~al.}}]{kajiwara2010transmission}%
  \BibitemOpen
  \bibfield  {author} {\bibinfo {author} {\bibfnamefont {Y.}~\bibnamefont
  {Kajiwara}}, \bibinfo {author} {\bibfnamefont {K.}~\bibnamefont {Harii}},
  \bibinfo {author} {\bibfnamefont {S.}~\bibnamefont {Takahashi}}, \bibinfo
  {author} {\bibfnamefont {J.}~\bibnamefont {Ohe}}, \bibinfo {author}
  {\bibfnamefont {K.}~\bibnamefont {Uchida}}, \bibinfo {author} {\bibfnamefont
  {M.}~\bibnamefont {Mizuguchi}}, \bibinfo {author} {\bibfnamefont
  {H.}~\bibnamefont {Umezawa}}, \bibinfo {author} {\bibfnamefont
  {H.}~\bibnamefont {Kawai}}, \bibinfo {author} {\bibfnamefont
  {K.}~\bibnamefont {Ando}}, \bibinfo {author} {\bibfnamefont {K.}~\bibnamefont
  {Takanashi}},  \emph {et~al.},\ }\href@noop {} {\bibfield  {journal}
  {\bibinfo  {journal} {Nature}\ }\textbf {\bibinfo {volume} {464}},\ \bibinfo
  {pages} {262} (\bibinfo {year} {2010})}\BibitemShut {NoStop}%
\bibitem [{\citenamefont {Sandweg}\ \emph {et~al.}(2010)\citenamefont
  {Sandweg}, \citenamefont {Kajiwara}, \citenamefont {Ando}, \citenamefont
  {Saitoh},\ and\ \citenamefont {Hillebrands}}]{sandweg2010enhancement}%
  \BibitemOpen
  \bibfield  {author} {\bibinfo {author} {\bibfnamefont {C.}~\bibnamefont
  {Sandweg}}, \bibinfo {author} {\bibfnamefont {Y.}~\bibnamefont {Kajiwara}},
  \bibinfo {author} {\bibfnamefont {K.}~\bibnamefont {Ando}}, \bibinfo {author}
  {\bibfnamefont {E.}~\bibnamefont {Saitoh}}, \ and\ \bibinfo {author}
  {\bibfnamefont {B.}~\bibnamefont {Hillebrands}},\ }\href@noop {} {\bibfield
  {journal} {\bibinfo  {journal} {Appl. Phys. Lett.}\ }\textbf {\bibinfo
  {volume} {97}},\ \bibinfo {pages} {252504} (\bibinfo {year}
  {2010})}\BibitemShut {NoStop}%
\bibitem [{\citenamefont {Liu}\ \emph {et~al.}(2011)\citenamefont {Liu},
  \citenamefont {Moriyama}, \citenamefont {Ralph},\ and\ \citenamefont
  {Buhrman}}]{liu2011spin}%
  \BibitemOpen
  \bibfield  {author} {\bibinfo {author} {\bibfnamefont {L.}~\bibnamefont
  {Liu}}, \bibinfo {author} {\bibfnamefont {T.}~\bibnamefont {Moriyama}},
  \bibinfo {author} {\bibfnamefont {D.}~\bibnamefont {Ralph}}, \ and\ \bibinfo
  {author} {\bibfnamefont {R.}~\bibnamefont {Buhrman}},\ }\href@noop {}
  {\bibfield  {journal} {\bibinfo  {journal} {Phys. Rev. Lett.}\ }\textbf
  {\bibinfo {volume} {106}},\ \bibinfo {pages} {036601} (\bibinfo {year}
  {2011})}\BibitemShut {NoStop}%
\bibitem [{\citenamefont {Azevedo}\ \emph {et~al.}(2011)\citenamefont
  {Azevedo}, \citenamefont {Vilela-Leao}, \citenamefont
  {Rodr{\'\i}guez-Su{\'a}rez}, \citenamefont {Santos},\ and\ \citenamefont
  {Rezende}}]{azevedo2011spin}%
  \BibitemOpen
  \bibfield  {author} {\bibinfo {author} {\bibfnamefont {A.}~\bibnamefont
  {Azevedo}}, \bibinfo {author} {\bibfnamefont {L.}~\bibnamefont
  {Vilela-Leao}}, \bibinfo {author} {\bibfnamefont {R.}~\bibnamefont
  {Rodr{\'\i}guez-Su{\'a}rez}}, \bibinfo {author} {\bibfnamefont {A.~L.}\
  \bibnamefont {Santos}}, \ and\ \bibinfo {author} {\bibfnamefont
  {S.}~\bibnamefont {Rezende}},\ }\href@noop {} {\bibfield  {journal} {\bibinfo
   {journal} {Phys. Rev. B}\ }\textbf {\bibinfo {volume} {83}},\ \bibinfo
  {pages} {144402} (\bibinfo {year} {2011})}\BibitemShut {NoStop}%
\bibitem [{\citenamefont {Kleinlein}\ \emph {et~al.}(2014)\citenamefont
  {Kleinlein}, \citenamefont {Ocker},\ and\ \citenamefont
  {Schmidt}}]{kleinlein2014using}%
  \BibitemOpen
  \bibfield  {author} {\bibinfo {author} {\bibfnamefont {J.}~\bibnamefont
  {Kleinlein}}, \bibinfo {author} {\bibfnamefont {B.}~\bibnamefont {Ocker}}, \
  and\ \bibinfo {author} {\bibfnamefont {G.}~\bibnamefont {Schmidt}},\
  }\href@noop {} {\bibfield  {journal} {\bibinfo  {journal} {Appl. Phys.
  Lett.}\ }\textbf {\bibinfo {volume} {104}},\ \bibinfo {pages} {153507}
  (\bibinfo {year} {2014})}\BibitemShut {NoStop}%
\bibitem [{\citenamefont {Nozaki}\ \emph {et~al.}(2012)\citenamefont {Nozaki},
  \citenamefont {Shiota}, \citenamefont {Miwa}, \citenamefont {Murakami},
  \citenamefont {Bonell}, \citenamefont {Ishibashi}, \citenamefont {Kubota},
  \citenamefont {Yakushiji}, \citenamefont {Saruya}, \citenamefont {Fukushima}
  \emph {et~al.}}]{nozaki2012electric}%
  \BibitemOpen
  \bibfield  {author} {\bibinfo {author} {\bibfnamefont {T.}~\bibnamefont
  {Nozaki}}, \bibinfo {author} {\bibfnamefont {Y.}~\bibnamefont {Shiota}},
  \bibinfo {author} {\bibfnamefont {S.}~\bibnamefont {Miwa}}, \bibinfo {author}
  {\bibfnamefont {S.}~\bibnamefont {Murakami}}, \bibinfo {author}
  {\bibfnamefont {F.}~\bibnamefont {Bonell}}, \bibinfo {author} {\bibfnamefont
  {S.}~\bibnamefont {Ishibashi}}, \bibinfo {author} {\bibfnamefont
  {H.}~\bibnamefont {Kubota}}, \bibinfo {author} {\bibfnamefont
  {K.}~\bibnamefont {Yakushiji}}, \bibinfo {author} {\bibfnamefont
  {T.}~\bibnamefont {Saruya}}, \bibinfo {author} {\bibfnamefont
  {A.}~\bibnamefont {Fukushima}},  \emph {et~al.},\ }\href@noop {} {\bibfield
  {journal} {\bibinfo  {journal} {Nat. Phys.}\ }\textbf {\bibinfo {volume}
  {8}},\ \bibinfo {pages} {491} (\bibinfo {year} {2012})}\BibitemShut {NoStop}%
\bibitem [{\citenamefont {Thiaville}\ and\ \citenamefont
  {Nakatani}(2008)}]{thiaville2008electrical}%
  \BibitemOpen
  \bibfield  {author} {\bibinfo {author} {\bibfnamefont {A.}~\bibnamefont
  {Thiaville}}\ and\ \bibinfo {author} {\bibfnamefont {Y.}~\bibnamefont
  {Nakatani}},\ }\href@noop {} {\bibfield  {journal} {\bibinfo  {journal} {J.
  App. Phys.}\ }\textbf {\bibinfo {volume} {104}},\ \bibinfo {pages} {093701}
  (\bibinfo {year} {2008})}\BibitemShut {NoStop}%
\bibitem [{\citenamefont {Griffiths}(2013)}]{griffiths2013introduction}%
  \BibitemOpen
  \bibfield  {author} {\bibinfo {author} {\bibfnamefont {D.}~\bibnamefont
  {Griffiths}},\ }\href@noop {} {\emph {\bibinfo {title} {Introduction to
  Electrodynamics}}}\ (\bibinfo  {publisher} {Pearson},\ \bibinfo {year}
  {2013})\BibitemShut {NoStop}%
\bibitem [{\citenamefont {Weisstein}(2004)}]{weisstein2004spherical}%
  \BibitemOpen
  \bibfield  {author} {\bibinfo {author} {\bibfnamefont {E.}~\bibnamefont
  {Weisstein}},\ }\href@noop {} {\bibfield  {journal} {\bibinfo  {journal}
  {MathWorld: The Web's Most Extensive Mathematics Ressource,[1999-2008]-[cit.
  16. 5. 2008]. Available online: http://mathworld. wolfram.
  com/SphericalCoordinates. html}\ } (\bibinfo {year} {2004})}\BibitemShut
  {NoStop}%
\bibitem [{\citenamefont {Dieny}\ \emph {et~al.}(1991)\citenamefont {Dieny},
  \citenamefont {Speriosu}, \citenamefont {Parkin}, \citenamefont {Gurney},
  \citenamefont {Wilhoit},\ and\ \citenamefont {Mauri}}]{dieny1991giant}%
  \BibitemOpen
  \bibfield  {author} {\bibinfo {author} {\bibfnamefont {B.}~\bibnamefont
  {Dieny}}, \bibinfo {author} {\bibfnamefont {V.~S.}\ \bibnamefont {Speriosu}},
  \bibinfo {author} {\bibfnamefont {S.~S.}\ \bibnamefont {Parkin}}, \bibinfo
  {author} {\bibfnamefont {B.~A.}\ \bibnamefont {Gurney}}, \bibinfo {author}
  {\bibfnamefont {D.~R.}\ \bibnamefont {Wilhoit}}, \ and\ \bibinfo {author}
  {\bibfnamefont {D.}~\bibnamefont {Mauri}},\ }\href@noop {} {\bibfield
  {journal} {\bibinfo  {journal} {Physical Review B}\ }\textbf {\bibinfo
  {volume} {43}},\ \bibinfo {pages} {1297} (\bibinfo {year}
  {1991})}\BibitemShut {NoStop}%
\bibitem [{\citenamefont {Parkin}\ \emph {et~al.}(1991)\citenamefont {Parkin},
  \citenamefont {Li},\ and\ \citenamefont {Smith}}]{parkin1991giant}%
  \BibitemOpen
  \bibfield  {author} {\bibinfo {author} {\bibfnamefont {S.}~\bibnamefont
  {Parkin}}, \bibinfo {author} {\bibfnamefont {Z.}~\bibnamefont {Li}}, \ and\
  \bibinfo {author} {\bibfnamefont {D.~J.}\ \bibnamefont {Smith}},\ }\href@noop
  {} {\bibfield  {journal} {\bibinfo  {journal} {Applied Physics Letters}\
  }\textbf {\bibinfo {volume} {58}},\ \bibinfo {pages} {2710} (\bibinfo {year}
  {1991})}\BibitemShut {NoStop}%
\bibitem [{\citenamefont {Donahue}\ and\ \citenamefont
  {Porter}(1999)}]{donahue1999oommf}%
  \BibitemOpen
  \bibfield  {author} {\bibinfo {author} {\bibfnamefont {M.}~\bibnamefont
  {Donahue}}\ and\ \bibinfo {author} {\bibfnamefont {D.}~\bibnamefont
  {Porter}},\ }\href@noop {} {\bibfield  {journal} {\bibinfo  {journal}
  {NISTIR}\ }\textbf {\bibinfo {volume} {6376}},\ \bibinfo {pages} {158}
  (\bibinfo {year} {1999})}\BibitemShut {NoStop}%
\bibitem [{\citenamefont {Sondheimer}(1952)}]{sondheimer1952mean}%
  \BibitemOpen
  \bibfield  {author} {\bibinfo {author} {\bibfnamefont {E.}~\bibnamefont
  {Sondheimer}},\ }\href@noop {} {\bibfield  {journal} {\bibinfo  {journal}
  {Adv. Phys.}\ }\textbf {\bibinfo {volume} {1}},\ \bibinfo {pages} {1}
  (\bibinfo {year} {1952})}\BibitemShut {NoStop}%
\bibitem [{\citenamefont {Desai}\ \emph {et~al.}(1984)\citenamefont {Desai},
  \citenamefont {Chu}, \citenamefont {James},\ and\ \citenamefont
  {Ho}}]{desai1984electrical}%
  \BibitemOpen
  \bibfield  {author} {\bibinfo {author} {\bibfnamefont {P.}~\bibnamefont
  {Desai}}, \bibinfo {author} {\bibfnamefont {T.}~\bibnamefont {Chu}}, \bibinfo
  {author} {\bibfnamefont {H.}~\bibnamefont {James}}, \ and\ \bibinfo {author}
  {\bibfnamefont {C.}~\bibnamefont {Ho}},\ }\href@noop {} {\bibfield  {journal}
  {\bibinfo  {journal} {J. Phys. Chem. Ref. Data}\ }\textbf {\bibinfo {volume}
  {13}},\ \bibinfo {pages} {1069} (\bibinfo {year} {1984})}\BibitemShut
  {NoStop}%
\bibitem [{\citenamefont {Eid}\ \emph {et~al.}(2003)\citenamefont {Eid},
  \citenamefont {Pratt~Jr},\ and\ \citenamefont {Bass}}]{eid2003enhancing}%
  \BibitemOpen
  \bibfield  {author} {\bibinfo {author} {\bibfnamefont {K.}~\bibnamefont
  {Eid}}, \bibinfo {author} {\bibfnamefont {W.}~\bibnamefont {Pratt~Jr}}, \
  and\ \bibinfo {author} {\bibfnamefont {J.}~\bibnamefont {Bass}},\ }\href@noop
  {} {\bibfield  {journal} {\bibinfo  {journal} {J. App. Phys.}\ }\textbf
  {\bibinfo {volume} {93}},\ \bibinfo {pages} {3445} (\bibinfo {year}
  {2003})}\BibitemShut {NoStop}%
\bibitem [{\citenamefont {Eid}\ \emph {et~al.}(2002)\citenamefont {Eid},
  \citenamefont {Fonck}, \citenamefont {Darwish}, \citenamefont {Pratt~Jr},\
  and\ \citenamefont {Bass}}]{eid2002current}%
  \BibitemOpen
  \bibfield  {author} {\bibinfo {author} {\bibfnamefont {K.}~\bibnamefont
  {Eid}}, \bibinfo {author} {\bibfnamefont {R.}~\bibnamefont {Fonck}}, \bibinfo
  {author} {\bibfnamefont {M.}~\bibnamefont {Darwish}}, \bibinfo {author}
  {\bibfnamefont {W.}~\bibnamefont {Pratt~Jr}}, \ and\ \bibinfo {author}
  {\bibfnamefont {J.}~\bibnamefont {Bass}},\ }\href@noop {} {\bibfield
  {journal} {\bibinfo  {journal} {J. App. Phys.}\ }\textbf {\bibinfo {volume}
  {91}},\ \bibinfo {pages} {8102} (\bibinfo {year} {2002})}\BibitemShut
  {NoStop}%
\bibitem [{\citenamefont {Yuasa}\ \emph {et~al.}(2002)\citenamefont {Yuasa},
  \citenamefont {Yoshikawa}, \citenamefont {Kamiguchi}, \citenamefont
  {Iwasaki}, \citenamefont {Takagishi}, \citenamefont {Sahashi} \emph
  {et~al.}}]{yuasa2002output}%
  \BibitemOpen
  \bibfield  {author} {\bibinfo {author} {\bibfnamefont {H.}~\bibnamefont
  {Yuasa}}, \bibinfo {author} {\bibfnamefont {M.}~\bibnamefont {Yoshikawa}},
  \bibinfo {author} {\bibfnamefont {Y.}~\bibnamefont {Kamiguchi}}, \bibinfo
  {author} {\bibfnamefont {H.}~\bibnamefont {Iwasaki}}, \bibinfo {author}
  {\bibfnamefont {M.}~\bibnamefont {Takagishi}}, \bibinfo {author}
  {\bibfnamefont {M.}~\bibnamefont {Sahashi}},  \emph {et~al.},\ }\href@noop {}
  {\bibfield  {journal} {\bibinfo  {journal} {J. App. Phys.}\ }\textbf
  {\bibinfo {volume} {92}},\ \bibinfo {pages} {2646} (\bibinfo {year}
  {2002})}\BibitemShut {NoStop}%
\bibitem [{\citenamefont {Strelkov}\ \emph {et~al.}(2003)\citenamefont
  {Strelkov}, \citenamefont {Vedyaev},\ and\ \citenamefont
  {Dieny}}]{strelkov2003extension}%
  \BibitemOpen
  \bibfield  {author} {\bibinfo {author} {\bibfnamefont {N.}~\bibnamefont
  {Strelkov}}, \bibinfo {author} {\bibfnamefont {A.}~\bibnamefont {Vedyaev}}, \
  and\ \bibinfo {author} {\bibfnamefont {B.}~\bibnamefont {Dieny}},\
  }\href@noop {} {\bibfield  {journal} {\bibinfo  {journal} {J. App. Phys.}\
  }\textbf {\bibinfo {volume} {94}},\ \bibinfo {pages} {3278} (\bibinfo {year}
  {2003})}\BibitemShut {NoStop}%
\bibitem [{\citenamefont {Aharoni}(1998)}]{aharoni1998demagnetizing}%
  \BibitemOpen
  \bibfield  {author} {\bibinfo {author} {\bibfnamefont {A.}~\bibnamefont
  {Aharoni}},\ }\href@noop {} {\bibfield  {journal} {\bibinfo  {journal} {J.
  App. Phys.}\ }\textbf {\bibinfo {volume} {83}},\ \bibinfo {pages} {3432}
  (\bibinfo {year} {1998})}\BibitemShut {NoStop}%
\end{thebibliography}%

\end{document}